\documentclass{emulateapj}
\usepackage{epsf}
\usepackage{graphicx}
\usepackage{float}
\usepackage{afterpage}
\usepackage{relsize}
\usepackage{amsmath}
\usepackage{epsfig}
\usepackage{enumitem}
\usepackage{subfigure}

\def\ba{\begin{eqnarray}}
\def\ea{\end{eqnarray}}
\def\etal{et al. }

\begin{document}


\title{Planet formation in binaries: dynamics of planetesimals perturbed by the eccentric protoplanetary disk and the secondary}

\author{Kedron Silsbee\altaffilmark{1} \& Roman R. Rafikov\altaffilmark{1} }
\altaffiltext{1}{Department of Astrophysical Sciences, 
Princeton University, Ivy Lane, Princeton, NJ 08540; 
ksilsbee@astro.princeton.edu}

\begin{abstract}
Detections of planets in eccentric, close (separations of $\sim 20$ AU) binary systems such as $\alpha$ Cen or $\gamma$ Cep provide an important test of planet formation theories. Gravitational perturbations from the companion are expected to excite high planetesimal eccentricities resulting in destruction, rather than growth, of objects with sizes of up to several hundred km in collisions of similar-size bodies. It was recently suggested that gravity of a massive axisymmetric gaseous disk in which planetesimals are embedded drives rapid precession of their orbits, suppressing eccentricity excitation. However, disks in binaries are themselves expected to be eccentric, leading to additional planetesimal excitation. Here we develop secular theory of eccentricity evolution for planetesimals perturbed by the gravity of an elliptical protoplanetary disk (neglecting gas drag) and the companion. For the first time we derive an expression for the disturbing function due to an eccentric disk, which can be used for a variety of other astrophysical problems. We obtain explicit analytical solutions for planetesimal eccentricity evolution neglecting gas drag and delineate four different regimes of dynamical excitation. We show that in systems with massive ($\gtrsim 10^{-2}M_\odot$) disks, planetesimal eccentricity is usually determined by the gravity of the eccentric disk alone, and is comparable to the disk eccentricity. As a result, the latter imposes a lower limit on collisional velocities of solids, making their growth problematic. In the absence of gas drag this fragmentation barrier can be alleviated if the gaseous disk rapidly precesses or if its own self-gravity is efficient at lowering disk eccentricity.    
\end{abstract}



\section{Introduction}


Planet-hosting binary systems with separations of several tens of AU present an interesting testbed for planet formation theories. Strong gravitational perturbations induced by the companion excite high eccentricities of planetesimals out of which planets form. Agglomeration of these objects into bigger bodies in mutual collisions, most effective at low relative speeds because of gravitational focussing, may become very ineffective. In a strongly dynamically excited environment planetesimals would destroy each other instead of growing. This {\it fragmentation barrier} presents a very serious problem for planetary growth in binaries. 

This issue is particularly severe for binaries with small separation. At the moment, we know (Chauvin et al. 2011; Dumusque et al. 2012) of five planet-hosting systems with eccentric companions (eccentricities $\gtrsim 0.4$) and semimajor axes of about 20 AU.  Three of them --- HD196885, $\gamma$ Cep, HD 41004 --- harbor giant planets with masses above that of the Jupiter at $1.6-2.6$ AU. At these separations, the eccentricity of a free particle can easily reach $0.1$ (Heppenheimer 1978), leading to collisions at speeds of several km s$^{-1}$ and resulting in destruction of even rather massive (several hundred km in size) objects in collisions, as well as smaller planetesimals. Two other systems --- $\alpha$ Cen and Gl 86 --- harbor planets at $\lesssim 0.1$ AU but even these objects have likely formed further out and then migrated in. 

Planetesimal agglomeration must proceed in gaseous protoplanetary disks. It has long been recognized that gas drag is an important agent of planetesimal dynamics (Marzari \& Scholl 2000; Th\'ebault et al. 2004, 2006, 2008, 2009; Paardekooper et al. 2008), helping lower relative speeds of planetesimals to some extent. Recently it has also been realized that the gravitational field of a massive protoplanetary disk can have a strong effect on planetesimal dynamics. In particular, Rafikov (2013b, hereafter R13) has shown that an {\it axisymmetric}, massive gaseous disk drives fast precession of planetesimal orbits by its gravity, which effectively suppresses eccentricity excitation by the companion. This mechanism permits growth of even $10$ km planetesimals at 2 AU as long as the disk is massive ($\sim 0.1M_\odot$) and axisymmetric.

At the same time hydrodynamical simulations of protoplanetary disks in binaries always find that disks perturbed by the companion develop some degree of non-axisymmetry (Okazaki et al. 2002; Kley et al. 2008; Marzari et al. 2009; Paardekooper et al. 2008), which usually manifests itself as a non-zero disk eccentricity. Such a disk has a non-axisymmetric component of its gravitational field which affects planetesimals in a way similar to the binary companion. Thus, one expects an eccentric gaseous disk to drive planetesimal eccentricity {\it excitation} (in addition to that produced by the binary companion), an effect absent in the case of an axisymmetric disk studied in R13. Recent work of Marzari et al. (2013) supports this expectation by showing this effect to operate in {\it circumbinary} disks, which can also develop eccentric structure and drive eccentricity growth by their gravity.   

The goal of this work is to analyze dynamics of planetesimals in the presence of gravitational perturbations due to both the binary companion and the eccentric disk. To focus on purely gravitational effects we neglect gas drag in our calculations (it is taken into account in Rafikov \& Silsbee 2014a,b). We explore planetesimal dynamics in the secular approximation, neglecting short-period perturbations of planetesimal orbits that average out over the long time intervals. The majority of our results are derived for the case of a non-precessing disk, which is steady with respect to the orientation of the eccentric orbit of the secondary. However, we also explore planetesimal dynamics in the case of precessing disk.

A significant part of this work is a derivation of the disturbing function due to an eccentric disk, which has been carried out for the first time. Because of the technical nature of this derivation, which we cover in Appendix \ref{sect:dist_fun}, it can be skipped at first reading. The main results are summarized in the main text. 

The structure of the paper is as follows. We outline the problem set-up in \S \ref{sect:setup} and present basic equations of planetesimal motion and their solutions for the case of non-precessing disk in \S \ref{sect:eqs}. We analyze our solutions and describe four possible dynamical regimes for planetesimal eccentricity excitation in \S \ref{sect:ecc_behavior}. Eccentricity behavior as a function of the distance from the primary is discussed in \S \ref{sect:ecc}.  In \S \ref{sect:precession} we explore the case of a uniformly precessing disk. Our results are discussed in \S \ref{sect:disc}, where we cover the implications for planetesimal growth (\S \ref{sect:planet_growth}), ways of lowering planetesimal eccentricity (\S \ref{sect:lowering_exc}), and comparison with existing numerical results (\S \ref{sect:num_compare}). Our findings are summarized in \S \ref{sect:sum}.


\section{Problem setup.}
\label{sect:setup}


We consider a binary star in which the primary and secondary have masses $M_p$ and $M_s$,   and define $\nu \equiv M_s/M_p$. The semimajor axis and eccentricity of the binary are $a_b$ and $e_b$, and its orientation is specified by apsidal angle $\varpi_b$. 

\begin{figure}
\hspace{-0.7cm}
\includegraphics[width=0.5\textwidth]{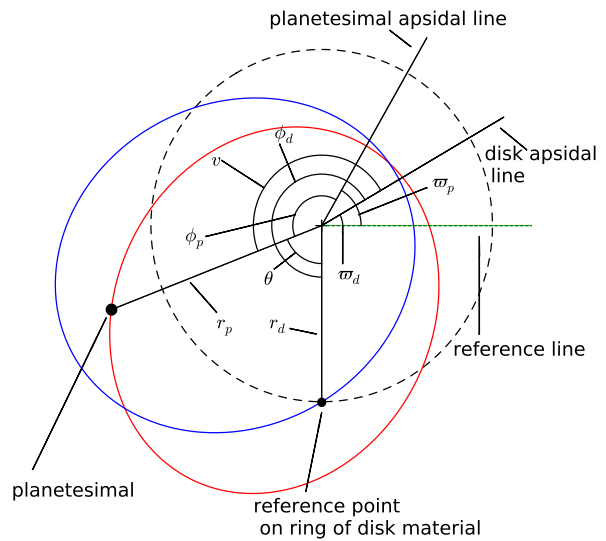}
\caption{Geometry of the problem, showing elliptical trajectories of both the planetesimal (red) and a representative fluid element (blue). Their orientation is shown using different polar angles. Dashed circle illustrates our calculation of the disturbing function in Appendix \ref{sect:dist_fun}.}
\label{fig:geometry}
\vspace{-.05cm}
\end{figure}

Coplanar with the binary and orbiting the primary star (this designation is arbitrary) is the {\it eccentric} gaseous disk with a non-axisymmetric surface density distribution $\Sigma(r_d,\phi_d)$. The disk is eccentric in a sense that {\it trajectories of its fluid elements are confocal ellipses}, which in general is not equivalent to $\Sigma$ being constant along these ellipses (see the discussion of this approximation in \S \ref{sect:disc}). We define $r_d$ to be the distance from the common focus of the elliptical fluid trajectories, and $\phi_d$ to be the polar angle with respect to the disk apsidal line, see Figure \ref{fig:geometry} for illustration. For every such gaseous trajectory with semimajor axis $a_d$ we can define the disk surface density at the periastron $\Sigma_p(a_d)$ and the eccentricity of the fluid trajectory $e_d(a_d)$, which we will simply call {\it disk eccentricity}. In general both $\Sigma_p(a_d)$ and $e_d(a_d)$ can be arbitrary functions of the fluid semi-major axis $a_d$, as long as $e_d(a_d)$ varies slowly enough for the particle trajectories to be non-crossing (Ogilvie 2001). 

Statler (1999) has given the following expression for the surface density behavior in such a disk, assuming that the lines of apsides of all elliptical trajectories are aligned:
\ba
\Sigma(a_d,\phi_d)=\Sigma_p(a_d)\frac{1-e_d^2-\zeta e_d (1+e_d)}
{1-e_d^2-\zeta e_d \left[e_d+\cos E(\phi_d)\right]},
\label{eq:Sigma}
\ea
where $\Sigma_p(a_d)$ is the surface density at the pericenter ($\phi_d=E=0$), as a function of the semi-major axis $a_d$, $E(\phi_d)$ is the eccentric anomaly (Murray \& Dermott 1999) and $\zeta \equiv d\ln e_d(a_d)/d\ln a_d$. Equation (\ref{eq:Sigma}) has been generalized in Statler (2001) and Ogilvie (2001) to the case of the disk apsidal angle $\varpi_d$ varying with $a_d$ but we will not consider this additional complication here as it adds little new to the physics of our problem. Interestingly, equation (\ref{eq:Sigma}) predicts that surface density is constant along the elliptical fluid trajectory if $e_d$ is not varying with $a_d$, i.e. $\zeta=0$. 

Throughout this work we assume simple power law scalings 
\ba
\Sigma_p(a_d)=\Sigma_0\left(\frac{a_{\rm out}}{a_d}\right)^p,~~~
e_d(a_d)=e_0\left(\frac{a_{\rm out}}{a_d}\right)^q,
\label{eq:Sigma0}
\ea
for $a_{\rm in}<a_d<a_{\rm out}$, where $a_{\rm in}$ and $a_{\rm out}$ are the semi-major axes of the innermost and outermost fluid trajectories, and $\Sigma_0$ and $e_0$ are the pericenter surface density and eccentricity {\it at the outer edge of the disk}. If the semi-major axis of the innermost fluid trajectory $a_{\rm in}\ll a_{\rm out}$, as expected for realistic disks, then $\Sigma_0$ can be directly related to the disk mass $M_d\approx 2\pi \int^{a_{\rm out}}_{a_{\rm in}}\Sigma_p(a_d)a_d da_d$ enclosed within $a_{\rm out}$ as
\ba
\Sigma_0=\frac{2-p}{2\pi}\frac{M_d}{a_{\rm out}^2},
\label{eq:sig_0}
\ea
where we neglected disk ellipticity (see below) and assumed $p<2$, so that most of the disk mass is concentrated in its outer part.

We will neglect the precession of the binary apsidal line caused by the gravity of the circumprimary disk, as the corresponding precession period is considerably longer than other timescales of the problem. We will also focus predominantly on the case of a non-precessing disk. We cover the precessing disk case in Appendix \ref{sect:prec_App} and \S \ref{sect:precession}.

Our focus is on the dynamics of planetesimals embedded in the gaseous disk. We characterize planetesimal orbits by semimajor axis $a_p$, eccentricity $e_p$, and apsidal angle $\varpi_p$. 

Even though expression (\ref{eq:Sigma}) does not assume $e_d$ to be small, in the rest of the paper we will take both the disk and planetesimal eccentricities to be small, $e_d(r)\ll 1$ and $e_b\ll 1$.


\section{Basic equations}
\label{sect:eqs}


We study planetesimal dynamics taking into account gravitational perturbations from both the binary companion and the eccentric disk. We perform calculations in the secular approximation (Murray \& Dermott 1999), by averaging the planetesimal disturbing function $R$ over time thus eliminating the short-period terms, and keeping only the slowly varying contributions up to second order in the planetesimal eccentricity $e_p$ and to lowest order in disk eccentricity $e_d$ (in all terms).


\subsection{Disturbing function due to the disk}
\label{sect:dist_disk}

In Appendix \ref{sect:dist_fun} we provide a detailed calculation of the planetesimal disturbing function $R_d$ due to a non-axisymmetric disk with surface density and eccentricity distributions given by equations (\ref{eq:Sigma0}). This calculation is very general and can be applied to an arbitrary eccentric disk, not necessarily around one of the components of the binary. In particular it can be used to study planetesimal motion in a circumbinary disk. This calculation thus represents an important stand-alone result of this work. 

We show in Appendix \ref{sect:dist_fun} that in the secular approximation and to lowest order in $e_d$ and $e_p$ the disturbing function due to the eccentric disk with orientation $\varpi_d$ (independent of the distance from the primary) has the form
\begin{equation}
R_d = a_p^2n_p \left[\frac{1}{2}A_de_p^2 + B_de_p \cos{(\varpi_p - \varpi_d)}\right],
\label{eq:R_d}
\end{equation}
where 
\ba
A_d & = & 2\pi\frac{G \Sigma_p(a_p)}{a_p n_p}\psi_1,
\label{eq:A_d}\\
B_d & = & \pi \frac{G \Sigma_p(a_p)}{a_p n_p}e_d(a_p)\psi_2,
\label{eq:B_d}
\ea
where $n_p\equiv \sqrt{GM_p/a_p^3}$ is the planetesimal mean motion, and dimensionless constants $\psi_1$ and $\psi_2$ are given by equations (\ref{psi1}) and (\ref{psi2}). In deriving this expression for $R_d$ we used equation (\ref{eq:dist_f}), in which we dropped the term independent of $e_p$.

\begin{figure}
\centering
\includegraphics[width=0.5\textwidth]{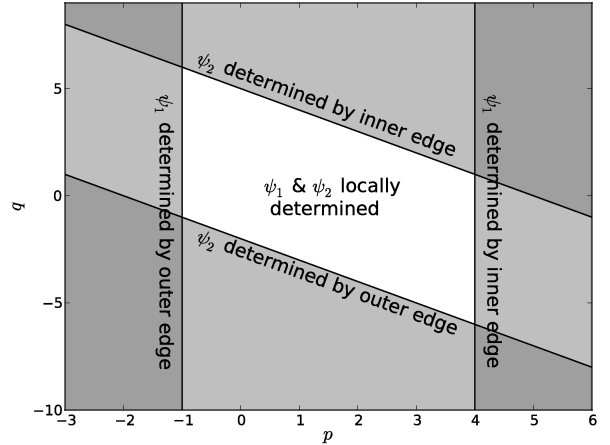}
\caption{Illustration of the convergence properties of coefficients $\psi_1$ and $\psi_2$ characterizing disk-driven precession and eccentricity excitation [equations (\ref{eq:A_d}) and (\ref{eq:B_d})] as a function of the power law indices $p$ and $q$ determining the radial dependence of disk surface density and eccentricity (equation \ref{eq:Sigma0}). The unshaded region is a part of parameter space where (far from the edges of the disk) the values of $\psi_1$ and $\psi_2$ are determined by the local disk properties at each radius, and is described by the constraint (\ref{eq:conditions}). Outside of this region the boundary terms must be accounted for in all of the disk, see Appendix \ref{sect:dist_fun} and Figure \ref{fig:psi_alpha}.}
\label{fig:pl_phase}
\vspace{-.05cm}
\end{figure}

\begin{figure}
\centering
\includegraphics[width=0.5\textwidth]{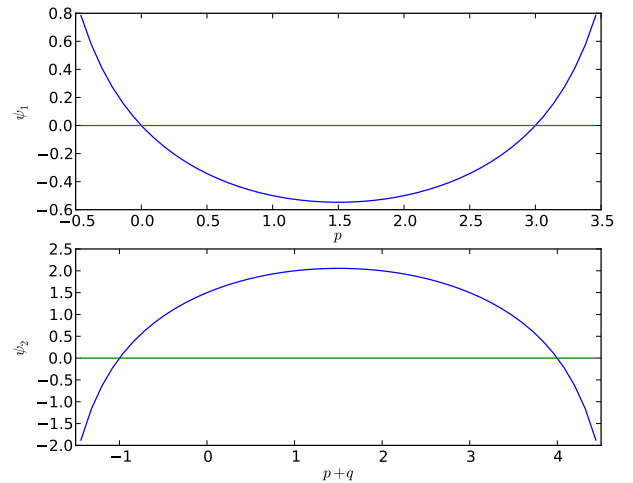}
\caption{Dependence of the coefficients (a) $\psi_1$ and (b) $\psi_2$ on power law indices $p$ and $p+q$, correspondingly (blue line). Calculation assumes that conditions (\ref{eq:conditions}) are fulfilled (unshaded region in Figure \ref{fig:pl_phase}) so that values of $\psi_{1,2}$ are determined by the local disk properties at each radius.}
\label{fig:psi_pl}
\vspace{-.05cm}
\end{figure}

Coefficients $\psi_1$ and $\psi_2$ are functions of the power law indices $p$, $q$, characterizing the disk structure, as well as the distance $a_p$ with respect to the disk boundaries. Figure \ref{fig:psi_alpha} shows the behavior of $\psi_1$ and $\psi_2$ for several values of $p$, $q$, and different $\alpha_1\equiv a_{\rm in}/a_p\le 1$, $\alpha_2\equiv a_p/a_{\rm out}\le 1$ computed according to equations (\ref{psi1})-(\ref{psi2}).  One can see that for the selected values of $p$ and $q$, both $\psi_1$ and $\psi_2$ converge to values depending only on $p$ and $q$ in the limit of $\alpha_1\to 0, \alpha_2\to 0$. Indeed, in Appendix \ref{sect:dist_fun} we show that as long as  
\ba
-1<p < 4~~~~\mbox{and}~~~~-2 < p + q < 5
\label{eq:conditions}
\ea
the values of $\psi_1$ and $\psi_2$ are determined locally, by the surface density and $e_d$ behavior in the vicinity of $a_p$. In this case, for a disk spanning more than an about order of magnitude in radius and $a_{\rm in}\lesssim a_p\lesssim a_{\rm out}$ the gravitational effect of disk parts near the boundaries is not important. Then $\psi_1$ and $\psi_2$ only weakly depend on $\alpha_{1,2}$ and can be well approximated by equations (\ref{eq:psi1_as})-(\ref{eq:psi2_as}). Their values in this limit are shown in Figure \ref{fig:psi_pl} as functions of $p$ and $p+q$. This is how these coefficients will be often treated (i.e. as constants) in the following analysis, though as can be seen in Figure \ref{fig:psi_alpha}, this approximation breaks down near the boundaries of the disk.

We verified our analytical derivation of $R_d$ given by equations (\ref{eq:R_d})-(\ref{eq:B_d}) in several different ways. In particular, in the case of an axisymmetric disk $B_d=0$, we made sure that in this case $R_d$ coincides with the expressions derived in R13 for surface density profile with $p=1$ and in Rafikov (2013b) for arbitrary $p$, based on the results of Ward (1981). The accuracy of our results in the case of non-axisymmetric disk is verified by direct integration of particle motion discussed in \S \ref{sect:compare}.


\subsection{Disturbing function due to the binary}
\label{sect:dist_bin}

Another perturbation to the planetesimal motion is provided by the companion star. For an external binary companion this is given by (Murray \& Dermott 1999)
\begin{equation}
R_b = a_p^2n_p \left[\frac{1}{2}A_be_p^2 + 
B_b e_p \cos{(\varpi_p - \varpi_b)}\right],
\label{eqdocbinpot}
\end{equation}
where 
\ba
A_b & = & \frac{\nu}{4}n_p \alpha_b^2 b_{3/2}^{(1)}(\alpha_b)\approx 
\frac{3}{4}n_p\nu \left(\frac{a_p}{a_b}\right)^3,
\label{eq:A_b}\\
B_b & = & - \frac{\nu}{4}n_p\alpha_b^2 b_{3/2}^{(2)}(\alpha_b)e_b\approx 
-\frac{15}{16}n_p\nu \left(\frac{a_p}{a_b}\right)^4e_b.
\label{eq:B_b}
\ea
Here $\alpha_b\equiv a_p/a_b$ and 
\ba
b_{s}^{(j)}(\alpha) = \frac{1}{\pi} \int_0^{2\pi} \frac{\cos{(j\theta)} d\theta}{(1 - 2 \alpha \cos{\theta} + \alpha^2)^s}
\label{eq:Laplace}
\ea
stands for the standard Laplace coefficient. The approximate expressions assume $\alpha_b\ll 1$, which is a reasonable assumption. Equations (\ref{eq:A_b})-(\ref{eq:B_b}) are valid up to the leading order in $e_b\ll 1$, more accurate expressions can be found in Heppenheimer (1978) or R13.


\subsection{Full planetesimal disturbing function}
\label{sect:dist_full}

Given that the binary precession due to disk gravity is slow, the orientation of the orbital ellipse of the secondary can be approximated as fixed in time. Then, without loss of generality we may choose the binary apsidal line as the reference direction, in which case $\varpi_b = 0$. The total (disk plus star) disturbing function $R=R_d+R_b$ is then given by 
\ba
\label{distFunc}
R = a_p^2n_p \Big[\frac{1}{2}Ae_p^2 & + & B_d e_p \cos\left(\varpi_p-\varpi_d\right)
\nonumber  \\
& + & B_b e_p \cos{\varpi_p}\Big],
\ea
where 
\ba
A = A_d + A_b.
\label{eq:A}
\ea

We now introduce planetesimal eccentricity vector ${\bf e}_p=(k_p,h_p)$, where 
\ba
k_p = e_p \cos\varpi_p,~~~~~ h_p = e_p \sin\varpi_p.
\label{eq:hk}
\ea 
Then $R$ can be written in terms of $h_p$ and $k_p$ as follows:
\ba
R = a_p^2n_p\Big[\frac{1}{2}A(h_p^2 + k_p^2) & + &  \left(B_b + B_d\cos\varpi_d\right)k_p
\nonumber  \\
& + &  B_d \sin\varpi_d h_p\Big].
\label{eq:R}
\ea


\subsection{Evolution equations and their solution}
\label{sect:ev_eq}

In secular planar approximation only the eccentricity $e_p$ and apsidal angle $\varpi_p$ of the planetesimal orbit vary in time. We study this process by following the evolution of $k_p$ and $h_p$ using Lagrange equations (Murray \& Dermott 1999)
\begin{equation}
\frac{dk_p}{dt} = -\frac{1}{n_pa_p^2} \frac{\partial R}{\partial h_p},
~~~~~~\frac{dh_p}{dt} = \frac{1}{n_pa_p^2} \frac{\partial R}{\partial k_p}.
\label{eq:lagrange}
\end{equation}
With $R$ given by the expression (\ref{eq:R}) the evolution equations become
\ba
\frac{dk_p}{dt} & = & -Ah_p - B_d \sin\varpi_d.
\label{eq:k_ev}\\
\frac{dh_p}{dt} & = & Ak_p + B_b + B_d \cos\varpi_d,
\label{eq:h_ev}
\ea
This is the key system of equations for our work, valid as long as the orientation of elliptical fluid trajectories, determined by $\varpi_d$, is independent of radius. 

Note that in deriving this system we did not make any assumptions regarding the time behavior of $\varpi_d$. Thus, $\varpi_d$ in equations (\ref{eq:k_ev})-(\ref{eq:h_ev}) can be an arbitrary function of time, which makes this system of equations applicable to rigidly precessing disks as well as disks in which the common apsidal line librates around some equilibrium orientation. 

However, for simplicity we start with a case when $\varpi_d=$ const, i.e. when the disk shape is fixed in the frame of the binary. The precessing disk case is covered in \S \ref{sect:precession}. We solve equations (\ref{eq:k_ev})-(\ref{eq:h_ev}) assuming an initially circular planetesimal orbit, i.e. $k_p(0) = h_p(0) = 0$. The solution 
\ba
\left\{
\begin{array}{l}
k_p(t)\\
h_p(t)
\end{array}
\right\} & = & {\bf e}_p(t)
= {\bf e}_{{\rm forced},b}+
{\bf e}_{{\rm forced},d}+{\bf e}_{\rm free},
\label{eq:decompose}\\
{\bf e}_{{\rm forced},b} & = & 
-\frac{B_b}{A}
\left\{
\begin{array}{l}
1 \\
0
\end{array}
\right\},
\label{eq:e_forced_b}\\
{\bf e}_{{\rm forced},d} & = & 
-\frac{B_d}{A}
\left\{
\begin{array}{l}
\cos\varpi_d \\
\sin\varpi_d
\end{array}
\right\},
\label{eq:e_forced_d}\\
{\bf e}_{\rm free}(t) & = & A^{-1}\sqrt{B_d^2 + 2B_dB_b\cos{\varpi_d} + B_b^2}
\nonumber\\
&\times &
\left\{
\begin{array}{l}
\cos (At + \phi) \\
\sin (At + \phi)
\end{array}
\right\},
\label{eq:e_free}
\ea
is decomposed into three distinct contributions: ${\bf e}_{{\rm forced},b}$ is the forced eccentricity due to binary potential, ${\bf e}_{{\rm forced},d}$ is the forced eccentricity due to disk potential, and ${\bf e}_{\rm free}(t)$ is the free eccentricity vector rotating at the precession rate $A$, with the phase $\phi$ given by equation 
\ba
\sin\phi=\frac{B_d\sin\varpi_d}{\sqrt{B_d^2 + 2B_dB_b\cos{\varpi_d} + B_b^2}}.
\label{eq:phase}
\ea

Variation of eccentricity $e_p=\left(h_p^2+k_p^2\right)^{1/2}$ is given by a simple formula
\ba
e_p(t) = \frac{2}{A}\left|\sin\frac{At}{2}\right| 
\sqrt{B_d^2 + 2B_dB_b\cos{\varpi_d} + B_b^2}.
\label{eq:e_p}
\ea
This result shows that the maximum eccentricity ranges between $2|(|B_d| - |B_b|)/A|$ and $2|(|B_d| + |B_b|)/A|$ depending on the value of $\varpi_d$.  

For the subsequent discussion we will be using a characteristic eccentricity of
\begin{equation}
\label{echar}
e_{\rm{char}} = 2 \frac{|B_b| + |B_d|}{|A|}, 
\end{equation}
which is an upper bound on the $e_p$. This estimate ignores the dependence of $e_p$ on $\varpi_d$ and overlooks some interesting cases when $e_p$ can be significantly lower than $e_{\rm{char}}$, e.g. when 
\ba
|B_b| \approx |B_d|~~~ \mbox{and}~~~ \cos\varpi_d \approx -\mbox{sgn}\left(B_dB_b\right),
\label{eq:low_e}
\ea
(sgn$(z)$ is a sign function) a possibility that is discussed in more detail in \S \ref{sect:orient}.


\subsection{Comparison with direct orbit integrations}
\label{sect:compare}

To test our analytical prescription (\ref{eq:R_d})-(\ref{eq:B_d}) for the disk disturbing function $R_d$, we compared our theory with the results of direct numerical integration of planetesimal motion in the gravitational field of an eccentric disk. We consider a disk extending from $a_{\rm in} = 0.1$ AU to $a_{\rm out} = 5$ AU and having $\Sigma_p(1 {\rm AU}) =$ 100 g cm$^{-2}$. To isolate effects of the disk gravity we set the mass of the secondary to zero. The details of our numerical calculations are described in \S \ref{ap:num_ver}.

\begin{figure}
\centering
\includegraphics[width=0.54\textwidth]{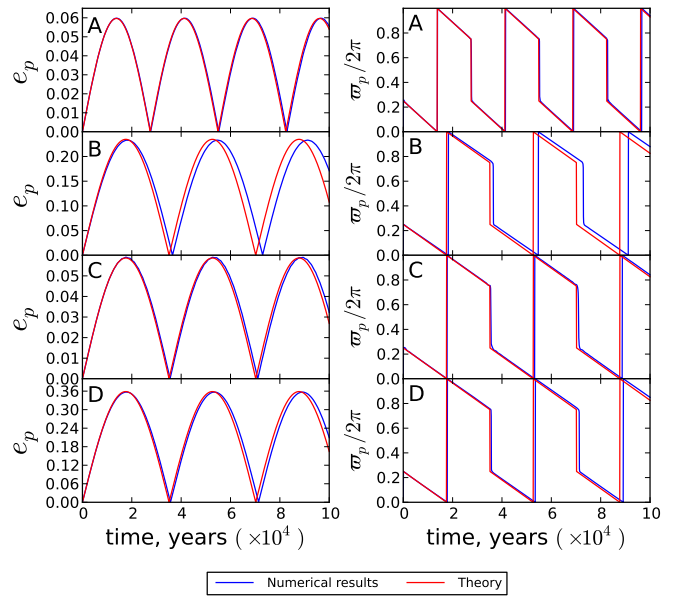}
\caption{Verification of analytical calculation of the disk disturbing function $R_d$ using numerical integrations with MERCURY. Time evolution of planetesimal eccentricity $e_p$ (left panels) and apsidal angle $\varpi_p$ (right panels) is shown for different disk parameters. Blue and red curves represent numerical and analytical results. In all cases planetesimals start with zero eccentricity, which explains the discontinuous jumps in $\varpi_p$: each time the orbit passes through zero eccentricity, $\varpi_p$ changes by $\pi$. The disk extends from 0.1 AU to 5 AU and has $\Sigma_p($1 AU) = 100 g cm$^{-2}$. (a) Planetesimal motion is shown at $a_p=$1 AU for a disk with $p = -q = 1$ and eccentricity at its outer edge $e_0 = 0.1$. (b) Same as (a) but at $a_p=$2 AU and $e_0$ = 0.2. (c) Same as (b) except that here eccentricity is lowered to $e_0 = 0.05$. (d) Here $a_p=$2 AU, $e_0 = 0.1$, and $p = 1$, but the disk eccentricity profile now has $q = 0$ --- $e_d$ is independent of distance. Apparently, in all cases the agreement between analytical secular theory and direct orbit integrations is very good. See text for more details. }
\label{fig:Mercury}
\vspace{-.05cm}
\end{figure}

Numerical results were then compared with analytical solutions obtained in the previous section, and the outcomes are shown in Figure \ref{fig:Mercury} in the form of planetesimal eccentricity $e_p$ and apsidal angle $\varpi_p$ dependence on time. We tried different initial conditions for planetresimal orbit but in this Figure we concentrate on the case of zero initial planetesimal eccentricity, when analytical solution is given by equation (\ref{eq:e_p}) with $B_b=0$.

One can see that irrespective of the parameters of our integrations the agreement between theory and numerical results is very good. The amplitude of $e_p$ variation is always in excellent agreement with theory (the difference being less than a percent), even for the disk eccentricity at the outer edge as high as $e_0=0.2$, see panel (b). The period of secular oscillations is within several percent of our analytical prediction $2\pi/A_d$ given by equation (\ref{eq:A_d}) in the high-eccentricity case $e_0=0.2$ for a disk with $p=-q=1$. However, this discrepancy is considerably smaller in other cases shown. 

Such deviations between orbit integrations and linear secular theory (although at much larger amplitude), predominantly in periodicity of variation, have been previously documented in the case of perturbation by the eccentric binary companion alone (Th\'ebault \etal 2006; Barnes \& Greenberg 2006). Giuppone \etal (2011) find discrepancies in both the amplitude and period of $e_p$ oscillations at the level of $\sim 50\%$ when secular theory predicts $e_p\gtrsim 0.1$.  But, as Figure \ref{fig:Mercury}b,d clearly demonstrates, the agreement between theory and simulations in the case of a disk is much better even when $e_p$ is as high as $0.2-0.4$. Most likely this is because the smooth mass distribution of the disk reduces the amplitude of its higher-order gravitational multipoles and allows secular theory, which goes only to octupole order, to better capture the main effects of the disk gravity.    

The general conclusion one can draw from the comparisons shown in Figure \ref{fig:Mercury} is that the secular theory for perturbations due to the disk developed in Appendix \ref{sect:dist_fun} works very well and our analytical results (\ref{psi1})-(\ref{psi2}) for the behavior of coefficients $\psi_1$ and $\psi_2$ are correct.


\section{Planetesimal eccentricity behavior}
\label{sect:ecc_behavior}


We will now consider different regimes of planetesimal dynamics. We start by using equation (\ref{eq:sig_0}) to express the disk-related precession rate $A_d$ and eccentricity excitation coefficient $B_d$ via the disk mass $M_d$:
\ba
A_d & = & (2-p)\psi_1 n_p \frac{M_d}{M_p}\left(\frac{a_p}{a_{\rm{out}}}\right)^{2-p},
\label{eq:A_d0}\\
B_d & = & \frac{2-p}{2}\psi_2 n_p \frac{M_d}{M_p}\left(\frac{a_p}{a_{\rm{out}}}\right)^{2-p} e_d(a_p), 
\label{eq:B_d0}
\ea
see equations (\ref{eq:A_d})-(\ref{eq:B_d}) and $e_d(a_p)$ is given by equation (\ref{eq:Sigma0}).

These expressions show that disk-driven planetesimal eccentricity is determined, in part, by the values of power law indices $p$ and $q$. Unfortunately, these parameters are rather poorly known for real protoplanetary disks. Based on standard accretion disk theory R13 advocated the use of $p\approx 1$ for the circumstellar disks in binaries. However, this choice is subject to uncertainly in our knowledge of the radial behavior of the viscous $\alpha$-parameter, thermal structure of the disk, etc. Thus, in this work we explore a range of values of $p$.

Equally uncertain is the choice of the disk eccentricity slope $q$. If one were to neglect the self-gravity, pressure and viscous forces in the gaseous disk then its fluid elements would behave as free particles perturbed by the binary companion and have their eccentricity scaling linearly with $a_p$ (Heppenheimer 1978; also equation [\ref{eq:BB_ecc}]), $e_d\propto a_p$, so that $q=-1$. This behavior is at least approximately supported by the numerical results of Okazaki et al. (2002) and semi-analytical calculations of Paardekooper et al. (2008) within a range of radii. Other authors find $e_d$ to exhibit more complicated, non-power law behavior (Kley et al. 2008; Marzari et al. 2009). Despite that, in this work we will predominantly stick to using $q=-1$, but sometimes we will consider other values of $q<0$.

All disk models considered in this paper have eccentricity $e_d$ increasing with radius, and surface density decreasing with radius. Under these natural assumptions, the disk should dominate the motion of planetesimals close to the primary star, since $A_b$ and $B_b$ very rapidly grow with $a_p$ (while $A_d$ and $B_d$ can even decay with $a_p$ for certain values of $p$ and $q$).  Similarly, for less massive disks, in the outer parts of the disk the binary dominates both the precession and eccentricity excitation of planetesimals. Then we may ignore the disk-driven perturbations unless the binary orbit is completely circular, in which case eccentricity excitation is solely due to the gravity of elliptical disk.

\begin{figure}
\centering
\includegraphics[width=0.5\textwidth]{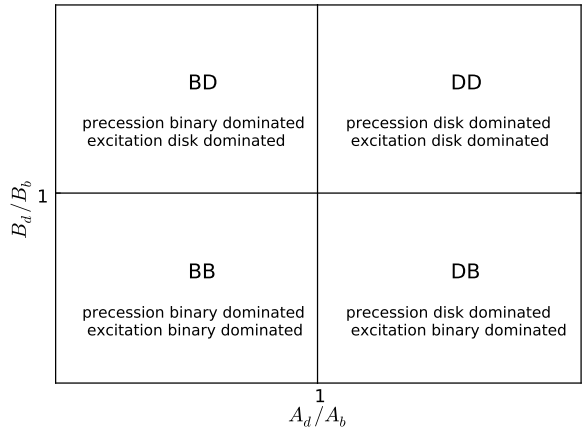}
\caption{
Illustration of different regimes of planetesimal eccentricity behavior, based on equations (\ref{AdAb}) and (\ref{BdBb}). Dynamical regimes are identified using two-letter notation as described in the text. See \S\S \ref{sect:DD}-\ref{sect:DB} for details.}
\label{fig:regimes}
\vspace{-.05cm}
\end{figure}

Using equations (\ref{eq:A_b}) and (\ref{eq:A_d0}) we can quantify this logic by forming a ratio
\ba
\left|\frac{A_d}{A_b}\right| & = & \frac{4|(2-p)\psi_1|}{3} \frac{M_d}{\nu M_p} \left(\frac{a_b}{a_{\rm{out}}}\right)^{2-p} 
\nonumber\\
& \times & \left(\frac{a_p}{a_b}\right)^{-(1+p)}.
\label{AdAb}
\ea
Disk (binary) terms dominate planetesimal precession rate when $|A_d/A_b|\gtrsim 1$ ($|A_d/A_b|\lesssim 1$), see Figure \ref{fig:regimes}. 

We do analogous calculation for eccentricity excitation using equations (\ref{eq:B_b}) and (\ref{eq:B_d0}):
\ba
\Big|\frac{B_d}{B_b}\Big| & = & \frac{8|(2-p)\psi_2|}{15} \frac{e_0}{e_b}\frac{M_d}{\nu M_p}\left(\frac{a_b}{a_{\rm{out}}}\right)^{2-p-q} 
\nonumber\\
& \times & \left(\frac{a_p}{a_b}\right)^{-(2+p+q)},
\label{BdBb}
\ea
where $e_0$ is the disk eccentricity at its outer edge. Again, disk (binary) dominates planetesimal eccentricity excitation when $|B_d/B_b|\gtrsim 1$ ($|B_d/B_b|\lesssim 1$), see Figure \ref{fig:regimes}. 

Conditions (\ref{AdAb}) \& (\ref{BdBb}) define special locations in the disk, where the ratios $|A_d/A_b|$, $|B_d/B_b|$ become equal to unity. We find that $|A_d/A_b|=1$ at
\ba
a_A & = & a_b\left[\frac{4|\psi_1 (2 - p)|}{3} \frac{M_d}{\nu M_p} 
\left(\frac{a_b}{a_{\rm out}}\right)^{2-p}\right]^{1/(1+p)}
\nonumber\\
& \approx & 0.16a_b\left[\frac{M_d/(\nu M_p)}{0.01}
~\frac{0.25}{a_{\rm out}/a_b}\right]^{0.5},
\label{eq:a_A}
\ea
while $|B_d/B_b|=1$ at
\ba
a_B & = & a_b\left[\frac{8|\psi_2(2-p)|}{15}\frac{M_d}{\nu M_p} \frac{e_{0}}{e_b}\left(\frac{a_b}{a_{\rm{out}}}\right)^{2-p-q}\right]^{1/(2+p+q)}
\nonumber\\
& \approx & 0.11a_b~\frac{0.25}{a_{\rm out}/a_b}
\left[\frac{M_d/(\nu M_p)}{0.01}
~\frac{e_0/e_b}{0.1}\right]^{0.5},
\label{eq:a_B}
\ea
where numerical estimates are for a disk model with $p=1$, $q=-1$ ($|\psi_1(1)|=0.5$, $|\psi_2(0)|=1.5$).

For the parameters adopted in these estimates both $a_A$ and $a_B$ lie within the disk, at separations of $2-3$ AU for $a_b=20$ AU (with $a_{\rm out}=5$ AU), which is outside the semi-major axes of the planets in binaries detected so far. The obvious implication is that these planets have formed in the part of the disk where secular effects were {\it dominated by the disk gravity} rather than by the secondary. This suggests that {\it disk gravity plays a decisive role} is determining planetesimal dynamics in the planet-building zone.  

Using ratios (\ref{AdAb}) \& (\ref{BdBb}) we now describe different possible regimes of the planetesimal eccentricity behavior, as illustrated in Figure \ref{fig:regimes}. We identify each regime using a two-letter notation in which the first letter describes what dominates planetesimal precession rate $A$, while the second refers to the dominance of eccentricity excitation (e.g. ``Case DB'' means that $|A_d/A_b|\gtrsim 1$ and $|B_d/B_b|\lesssim 1$). In Figure \ref{fig:phase_space} we map out these different dynamical regimes in the space of the scaled disk mass $M_d/(\nu M_p)$ and planetesimal semi-major $a_p/a_b$ for different disk models (combinations of $p$, $q$, $e_0/e_b$).


\subsection{Case DD: disk dominates both precession and excitation}
\label{sect:DD}

At small separations from the primary, $a_p\lesssim a_A, a_B$, the disk dominates both precession and eccentricity excitation of planetesimals, so $A\approx A_d$ and $|B_b|\ll |B_d|$. In this case the characteristic planetesimal eccentricity (\ref{echar}) tends to  
\ba
e_p^{\rm DD}(a_p)\to 2\left|\frac{B_d}{A_d}\right| = \left|\frac{\psi_2 }{\psi_1}\right|e_d(a_p).
\label{eq:DD_ecc}
\ea
In this regime the maximum planetesimal eccentricity is of order the local disk eccentricity, since $|\psi_{1,2}|\sim 1$. For example, ignoring edge effects $e_p(a_p)\to 3e_d(a_p)$ for a $p=1$, $q=-1$ disk.  Thus, an elliptical disk is capable of exciting planetesimal eccentricity of order of its own eccentricity $e_d$ purely by its non-axisymmetric gravitational field. In this regime planetesimal eccentricity should increase with $a_p$ because $e_d(a_p)$ is expected to be a growing function of $a_p$. 

Figure \ref{fig:phase_space} demonstrates that this dynamical regime is unavoidable for $a_p\lesssim 1$ AU even for relatively small disk masses, down to $M_d\sim 10^{-3}M_\odot$.


\subsection{Case BB: binary dominates both precession and excitation}
\label{sect:BB}

In the opposite limit, far from the primary, as $a_A, a_B\lesssim a_p$ (which is of course possible only if $a_A, a_B\lesssim a_{\rm out}$), planetesimal dynamics is governed completely by the binary potential.  The contribution from the disk is insignificant so that both $A\approx A_b$ and $|B_d|\ll |B_b|$. This is the limit of planetesimal dynamics in a diskless binary, which has been investigated by Heppenheimer (1978).

In this case planetesimal eccentricity is given by 
\ba
e_p^{\rm BB}(a_p)\to 2\left|\frac{B_b}{A_b}\right| = 
\frac{5}{2} \frac{a_p}{a_b} e_b,
\label{eq:BB_ecc}
\ea
in agreement with Heppenheimer (1978). 

Figure \ref{fig:phase_space} shows that Case BB is important for a broad range of separations, down to 1 AU, when the disk mass is very small, $\lesssim 10^{-3}M_\odot$. However, for more massive disks with $M_d\gtrsim 10^{-2}M_\odot$ this regime never emerges for $a_p<a_{\rm out}$. Thus, in compact binaries ($a_b\sim 20$ AU) with massive disks the classical result of Heppenheimer (1978) may never actually apply.


\subsection{Case BD: binary dominates precession, disk dominates excitation}
\label{sect:BD}

In between the two limiting cases covered in \S \ref{sect:DD} and \ref{sect:BB} there are other dynamical regimes.

Provided that $a_A<a_B$ there exists a region in the disk with $a_A\lesssim a_p \lesssim a_B$, where planetesimal precession is dominated by the binary companion ($A\approx A_b$), while eccentricity excitation is determined by the disk gravity ($|B_d|\gg |B_b|$). In this limit planetesimal eccentricity is given by 
\ba
e_p^{\rm BD}(a_p) & \to & 2\left|\frac{B_d}{A_b}\right| 
= \frac{4|\psi_2(2-p)|}{3} e_d(a_p)  
\frac{M_d}{\nu M_p} 
\nonumber\\
& \times &\left(\frac{a_b}{a_{\rm{out}}}\right)^{2-p}\left(\frac{a_p}{a_b}\right)^{-(1+p)}. 
\label{eq:BD_ecc}
\ea
Using this expression and equations (\ref{AdAb}), (\ref{BdBb}) one can easily show that 
\ba
e_p^{\rm BD}(a_p) & = & e_p^{\rm DD}(a_p)\left(\frac{a_p}{a_A}\right)^{-(1+p)}
\nonumber\\
& = & e_p^{\rm BB}(a_p)\left(\frac{a_p}{a_B}\right)^{-(2+p+q)}. 
\label{eq:BD_compare}
\ea
Since $a_A\lesssim a_p \lesssim a_B$ in Case BD, this result implies (for $p>-1$, $p+q>-2$) that $e_p^{\rm BB}(a_p)\lesssim e_p^{\rm BD}(a_p)\lesssim e_p^{\rm DD}(a_p)$.  It is then clear that Case BD requires $e_p^{\rm DD}(a_p)\gtrsim e_p^{\rm BB}(a_p)$ locally, i.e., according to equation (\ref{eq:DD_ecc}), that the disk eccentricity $e_d(a_p)$ be higher than planetesimal eccentricity $e_p^{\rm BB}$ in a diskless case for the values of $p$ and $q$ explored in this paper. This situation may not be easy to realize in practice since pressure and viscous forces may tend to reduce (and not increase) eccentricity of fluid elements compared to that expected for test particles (i.e. $e_p^{\rm BB}$).

\begin{figure}
\centering
\includegraphics[width=0.5\textwidth]{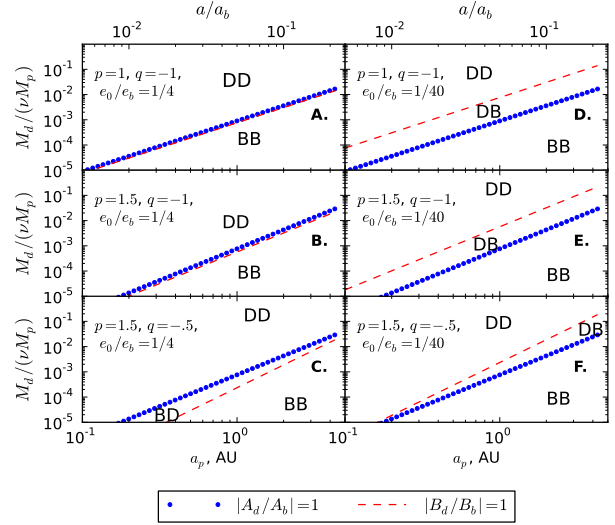}
\caption{Map of different dynamical regimes in the space of planetesimal semi-major axis $a_p$ and disk mass $M_d$. Different panels correspond to different disk models, ones on the right have disk eccentricity (indicated on panels together with $p$ and $q$) 10 times lower than the left ones. Disk extends from $0.1$ AU to 5 AU, binary semi-major axis $20$ AU, eccentricity $0.2$, and secondary to primary mass ratio $\nu=1/2$. Dynamical regimes in each part of the phase space are indicated. Dotted and dashed lines are given by equations (\ref{eq:a_A}) and (\ref{eq:a_B}). One can see that planetesimals are in DD regime in massive disks near the primary, and in BB regime in low-mass disks far from it. Edge effects are ignored in this calculation and  we use the values of $\psi_1$ and $\psi_2$ that they take at 1 AU.  See text for details.}
\label{fig:phase_space}
\vspace{-.05cm}
\end{figure}

Figure \ref{fig:phase_space} shows that indeed this dynamical regime requires rather special conditions to be realized, such as the relatively high value of the disk eccentricity $e_0/e_b$. Even then it typically occupies a narrow range of separations, see Figure \ref{fig:phase_space}a,b.  This is because disk models in these two panels have $e_d(a_p)\approx e_p^{\rm BB}(a_p)$, essentially eliminating Case BD region. In Figure \ref{fig:phase_space}c we do display a model with $e_d(a_p)\gtrsim e_p^{\rm BB}(a_p)$ close to the primary (we take $e_d\propto a_p^{1/2}$, while $e_p^{\rm BB}\propto a_p$) so that Case BD emerges at small $M_d$ and relatively small $a_p$. However, as we mentioned before, this may not be a typical situation.


\subsection{Case DB: disk dominates precession, binary dominates excitation}
\label{sect:DB}

Now we look at the opposite case of $a_B < a_A$, which emerges when $e_0/e_b$ is low. Within the range $a_B\lesssim a_p \lesssim a_A$ planetesimal precession is dominated by the disk gravity ($A\approx A_d$), while eccentricity excitation is determined predominantly by the secondary star ($|B_d|\ll |B_b|$). This is the approximation of a massive axisymmetric disk discussed in R13. In agreement with that work we find the maximum eccentricity to follow
\ba
e_p^{\rm DB}(a_p)\to 2\left|\frac{B_b}{A_d}\right| & = & 
\frac{15}{8|\psi_1(2-p)|}e_b\frac{\nu M_p}{M_d}
\nonumber\\
& \times & \left(\frac{a_{\rm out}}{a_b}\right)^{2-p}
\left(\frac{a_p}{a_b}\right)^{2+p}.
\label{eq:DB_ecc}
\ea
This expression and equations (\ref{AdAb}), (\ref{BdBb}) imply that 
\ba
e_p^{\rm DB}(a_p) & = & e_p^{\rm DD}(a_p)\left(\frac{a_p}{a_B}\right)^{2+p+q}
\nonumber\\
& = & e_p^{\rm BB}(a_p)\left(\frac{a_p}{a_A}\right)^{1+p}. 
\label{eq:DB_compare}
\ea
Because now $a_B\lesssim a_p \lesssim a_A$ we see that $e_p^{\rm DD}(a_p)\lesssim e_p^{\rm DB}(a_p)\lesssim e_p^{\rm BB}(a_p)$. Then it follows that DB regime requires $e_d(a_p)\sim e_p^{\rm DD}(a_p)\lesssim e_p^{\rm BB}(a_p)$ in non-pathological cases.

According to Figure \ref{fig:phase_space} this dynamical regime is rather common at low $e_0/e_b$, but is difficult to realize inside the disk for higher $e_0/e_b$. For some models (e.g. see Figure \ref{fig:phase_space}d,e) case DB regime holds within an extended region of the disk.

\begin{figure}
\centering
\includegraphics[scale = .35]{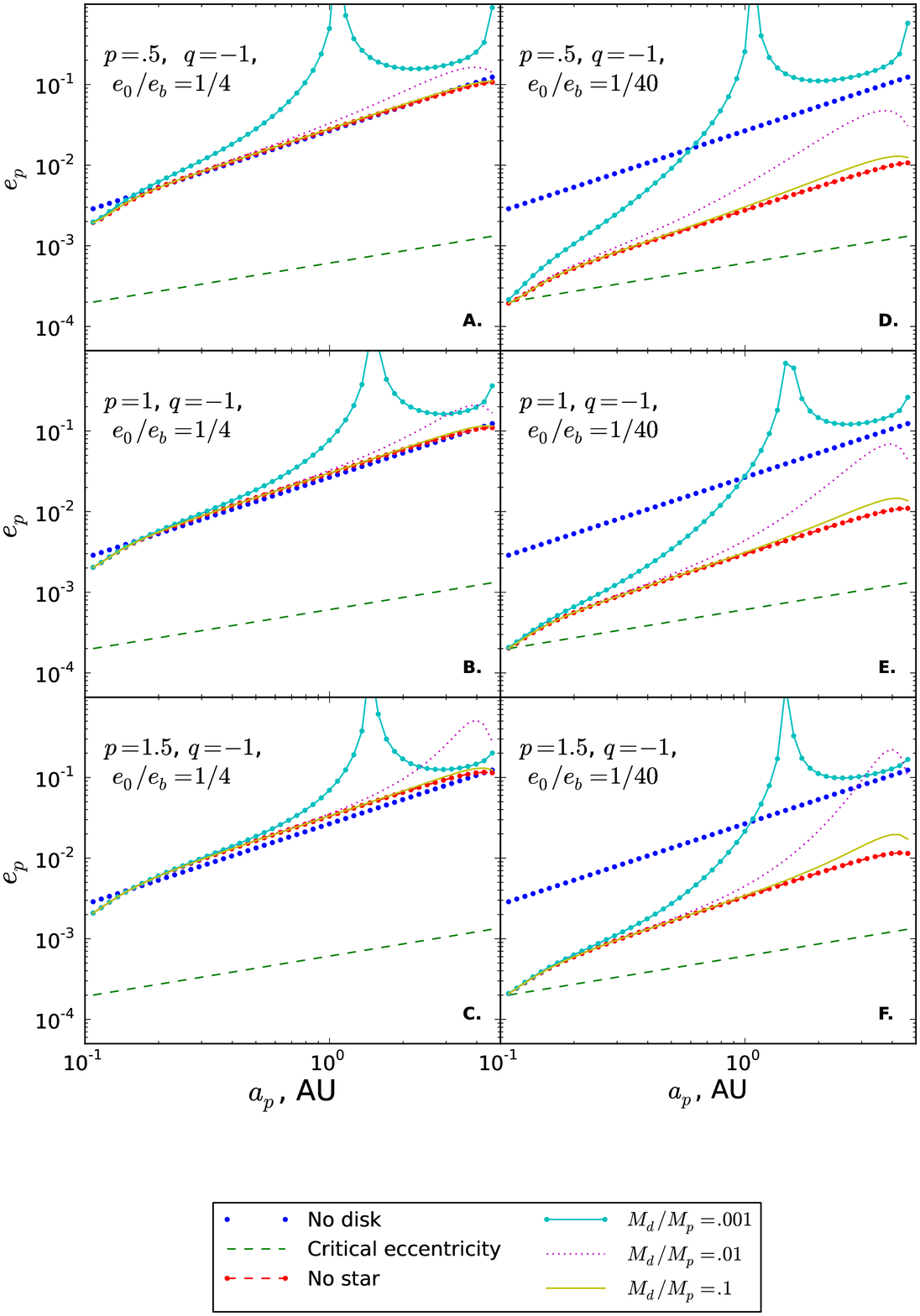}
\caption{Plots of planetesimal eccentricity as a function of $a_p$ for different disk models (values of $p$, $q$, and disk eccentricity at the outer disk edge $e_0$ are shown in panels).  For reference the dark blue big-dotted line shows eccentricity in the case of no disk (equation (\ref{eq:BB_ecc})), the red dot-dashed line shows the case of no secondary (disk only, equation (\ref{eq:DD_ecc})),  the green dashed line shows the critical eccentricity at fragmentation threshold (equation (\ref{ecrit})). Other curves show $e_p$ for a binary ($a_b=20$ AU, $e_b=0.2$, $M_p=M_\odot$, and $\nu=1/2$) with the disk extending from $0.1$ AU to 5 AU and having different mass as shown in panels. Note a conspicuous secular resonance around 1.5 AU in models with the low-mass disk. At small separations ($\lesssim 1$ AU) curves of $e_p(a_p)$ converge towards the disk-dominated solution, equation (\ref{eq:DD_ecc}). There are deviations of $e_p$ from simple power-law behavior at the inner and outer edges of the disk due to the nontrivial behavior of $\psi_1$ and $\psi_2$ there. See text for more details.}
\label{fig:ecc_r}
\vspace{-.05cm}
\end{figure}


\section{Eccentricity profiles}
\label{sect:ecc}


To illustrate results of the previous section, in Figure \ref{fig:ecc_r} we show profiles of planetesimal eccentricity computed for different disk models. For reference, each of the panels displays planetesimal eccentricity for the diskless case ($e_p^{\rm BB}(a_p)$, dark blue, big-dotted) as well as $e_p$ for the case with no secondary ($e_p^{\rm DD}(a_p)$, red dot-dashed). In the left panels we have chosen disk eccentricity $e_d(a_p)$ very close to the eccentricity of a free particle in the binary potential, which explains why the curves of $e_p^{\rm BB}(a_p)$ and $e_p^{\rm DD}(a_p)$ almost overlap. In the right panels $e_d$ is reduced by an order of magnitude and the two curves are well separated. 

Note that the $e_p^{\rm DD}(a_p)$ curve does not follow the simple power law in $a_p$ as one would have expected based on equation (\ref{eq:DD_ecc}) and the assumption of $\psi_1,\psi_2$ being constant --- it clearly deviates from this simple form at the disk edges. This is because near the disk edge, boundary terms neglected in computing the Figure \ref{fig:psi_pl} start to affect the values of $\psi_1$ and $\psi_2$ in a non-trivial manner, see Figure \ref{fig:psi_alpha}. 

We plot eccentricity profiles for different values of the disk mass. At the lowest disk mass, $M_d=10^{-3}M_p$, planetesimal eccentricity $e_p$ starts out very high in the outer disk (in the BB regime, see Figure \ref{fig:phase_space}), above $e_p^{\rm BB}(a_p)$. A notable feature of this profile is the {\it secular resonance} located at $\approx 1.5$ AU and causing $e_p$ to diverge. Its existence was predicted in R13 and Rafikov (2013a) for the case of circumprimary and circumbinary disks correspondingly. Later Meschiari (2014) confirmed the emergence of this resonance in massive circumbinary disks using numerical simulations of planetesimal dynamics. 

The origin of this resonance lies in the fact that $A_b$ is always positive, whereas for the disks that we are considering, $A_d$ is negative, see Figure \ref{fig:psi_pl}. This means that at $a_A$ (see equation (\ref{eq:a_A})), where $|A_d|=|A_b|$ one actually has $A=0$ and our secular solution (\ref{fig:psi_pl}) diverges. Inward of the resonance, $e_p$ rapidly goes down (in DB and DD regimes) and asymptotically approaches  $e_p^{\rm DD}(a_p)$ for $a_p\lesssim 0.5$ AU.

For a somewhat more massive disk $M_d=10^{-2} M_p$ the $e_p$ profile looks very different --- it does not exhibit secular resonance (since $a_A$ is now outside the outer disk edge $a_{\rm out}$) and generally features lower values of $e_p$.  This happens because with such a massive disk, planetesimal excitation is never in the BB regime. Disk gravity governs particle dynamics essentially through the whole disk.

This is even more so for the $M_d=10^{-1} M_p$ disk. At this high mass planetesimal eccentricity curves closely follow  $e_p^{\rm DD}(a_p)$ for all $a_p$.  As a result, in low-$e_0$ disks $e_p$ can be appreciably lower than what it is if planetesimals are affected by the gravity of the binary companion alone, similar to the case studied in R13.  Somewhat counerintuitively, adding an additional perturber --- a massive disk --- to the system does not heat it up dynamically but in fact {\it reduces} planetesimal random velocities.  

In all cases we see that $e_p$ is above the smaller of the $e_p^{\rm BB}(a_p)$ and $e_p^{\rm DD}(a_p)$. Thus, nonzero disk eccentricity introduces a {\it lower limit} on the $e_p$ value.


\section{Dynamics in the case of precessing disk}
\label{sect:precession}


So far we have been dealing with the case of non-precessing disk which keeps its orientation fixed in the frame of the binary orbit. However, simulations often find that gas disks in binaries not only develop a non-zero eccentricity but also precess (Okazaki et al. 2002; Paardekooper et al. 2008; Marzari et al. 2009). Thus it is important to discuss how planetesimal dynamics changes in the case of a precessing disk. 

In Appendix \ref{sect:prec_App} we present the extension of our solutions for the planetesimal eccentricity in \S \ref{sect:ev_eq} to the case of a disk that precesses as a solid body at a constant rate\footnote{Note that $\dot \varpi_d$ has a meaning different from that in R13, where $\dot \varpi_d$ was equivalent to $A_d$ in our current notation.} $\dot \varpi_d$. We find that the eccentricity vector can again be separated into three distinct contributions, see equation (\ref{eq:e_precess}): (1) standard forced eccentricity vector due to binary with amplitude $|{\bf e}_{{\rm forced},b}|=|B_b/A|$, stationary in the binary frame, (2) forced eccentricity vector due to the disk with amplitude $|{\bf e}_{{\rm forced},d}|=|B_d/(A-\dot\varpi_d)|$, rotating at the rate $\dot\varpi_d$, and (3) the free eccentricity term with amplitude
\ba
|{\bf e}_{\rm free}| & = & \frac{1}{|A(A-\dot \varpi_d)|}\Big[(AB_d)^2 + \left(B_b(A-\dot \varpi_d)\right)^2 
\nonumber\\
& + & 2AB_dB_b(A - \dot \varpi_d)\cos\varpi_{d0}\Big]^{1/2}
\label{eq:free}
\ea
rotating at the precession rate $A$ (here $\varpi_{d0}$ is the value of $\varpi_d$ at $t=0$).

The expression for the characteristic eccentricity becomes more complicated and depends on the value of $\varpi_{d0}$. The maximum possible eccentricity (for planetesimals starting with $h_p(0) = k_p(0) = 0$) is reached when $\varpi_d(0)=\varpi_{d0}=0$ or $\pi$ (disk and binary periapses aligned or anti-aligned initially), depending on the signs of $A$, $B_d$, and $A-\dot\varpi_d$. Then the maximum eccentricity is given by 
\begin{equation}
e_{\rm char}=2\left(\left|\frac{B_b}{A}\right|+
\left|\frac{B_d}{A-\dot\varpi_d}\right|\right).
\label{eq:echar_prec}
\end{equation}

Comparing this expression with equation (\ref{echar}) we conclude that disk precession does not affect planetesimal eccentricity behavior as long as $|\dot\varpi_d|\lesssim |A|$. 

However, in the opposite case of $|\dot\varpi_d|\gtrsim |A|$ the disk-driven forced part of the eccentricity vector is suppressed compared to the case of no precession. This is because rapid precession of the disk (compared to the rate of planetesimal orbital precession) effectively averages out the non-axisymmetric part of the disk potential, considerably reducing related eccentricity excitation. This has implications discussed in \S \ref{sect:rapid_precess}. At the same time the forced eccentricity contribution due to binary stays unchanged for planetesimals embedded in the precessing disk. We expect these asymptotic results to remain valid even in the case of non-uniform disk precession, both when it is much faster and much slower than $|A|$. However, all this discussion strictly applies only in the absence of gas drag.


\section{Discussion}
\label{sect:disc}


We can put our findings in the context of existing results on the purely gravitational dynamics (i.e. not accounting for gas drag) of planetesimals in binaries. Heppenheimer (1978) explored planetesimal dynamics under the gravity of the companion alone. Our results reduce to his in the limit of a zero-mass disk, i.e. when planetesimal dynamics is in the BB regime, see \S \ref{sect:BB}.

It was first shown analytically in R13 that the gravity of a massive disk can significantly suppress planetesimal eccentricity excitation in binaries. The reason lies in the fast precession of planetesimal orbits caused by the disk gravity, which effectively averages out $e_p$ forcing by the companion. This effect is present in our calculations as well and we reproduce the results of R13 in Case DB.

However, our work includes another important ingredient not considered previously in the framework of secular theory --- gravitational forcing of planetesimal eccentricity by the disk itself, which should be present in addition to planetesimal precession if the disk is eccentric. While some numerical studies on this topic do exist (see \S \ref{sect:num_compare}) analytical understanding of their results has been hampered by the complexity of the problem. 

In this work we have provided the first (to the best of our knowledge) calculation of the eccentric disk potential in application to planetesimal dynamics. Using this prescription we uncovered the existence of two entirely new regimes of planetesimal dynamics --- Case BD (\S \ref{sect:BD}) and Case DD (\S \ref{sect:DD}) --- in which eccentricity excitation by the disk exceeds that due to the secondary. The latter regime (DD) represents a very common situation in protoplanetary disks in binaries. As we have shown in \S \ref{sect:ecc_behavior} in many cases planetesimal excitation is in the DD regime throughout the whole disk. 

Significance of this dynamical regime also lies in the fact that the disk drives planetesimal eccentricities to a value of order the local disk eccentricity, see equation (\ref{eq:DD_ecc}). Even though in the absence of any damping agents eccentricity of a particle starting on a circular orbit oscillates, see equation (\ref{eq:e_p}), so that during some periods $e_p\ll e_{\rm char}$, most of the time  $e_p$ is of order $e_{\rm char}\sim e_d$ in the regime DD. Thus, eccentricity of the disk gives rise to a {\it lower limit} on the characteristic planetesimal eccentricity (\ref{echar}), which is a very important finding. 

In particular, it constrains the applicability of the axisymmetric disk approximation used in R13. Indeed, let us calculate $e_p$ using equation (\ref{eq:DB_ecc}), which is identical to the result of R13, for a system with $M_p=M_\odot$, $\nu=0.3$, $e_b=0.4$, $a_b=20$ AU harboring an axisymmetric disk with $a_{\rm out}=5$ AU, $M_d=10^{-2}M_\odot$, $p=1$. At $a_p=2$ AU we find $e_p\approx 10^{-2}$, which is much less that it would be in a diskless case, $e_p^{\rm BB}\approx 0.1$, see equation (\ref{eq:BB_ecc}). However, for this result to hold in a non-axisymmetric disk the disk eccentricity at 2 AU has to be less than $10^{-2}$. Whether such low $e_d$ is realistic is not clear at the moment (see \S \ref{sect:num_compare}).  

Our current results have been derived assuming that the disk affects planetesimals only via its gravitational field. In practice planetesimals are also subject to gas drag, which has important consequences for their dynamics. First, gas drag lowers planetesimal velocities with respect to gas, which also lowers relative planetesimal velocities therefore positively affecting survival in mutual collisions. Second, it has long been known that gas drag introduces apsidal alignment of planetesimal orbits (Marzari \& Scholl 2000), which considerably reduces relative collision velocities of near-equal bodies. However, planetesimals of different sizes would still collide at high speeds suppressing growth (Th\'ebault et al. 2006, 2008). Third, gas drag damps the free part of eccentricity, see Beaug\'e et al. (2010). This should affect the time dependence of planetesimal eccentricity, which in our case is given by equation (\ref{eq:e_p}). We address the effects of gas drag on planetesimal dynamics in binaries in Rafikov \& Silsbee (2014a). 

Because of the neglect of gas drag our current results are strictly valid only for relatively large objects, with sizes of several hundred km. For such planetesimals gas drag can be unimportant compared to purely gravitational forces during rather long time span, and may thus be neglected.  Inclusion of gas drag does not negate our finding that disk gravity from an eccentric disk leads to high encounter velocities between planetesimals, even of kilometer size.  Our results also clearly show that purely gravitational effects alone, in the absence of dissipative forces, can give rise to non-trivial behavior of $e_p$ (see e.g. \S \ref{sect:DD}, \ref{sect:BD}) not captured in previous analyses of the problem.

We also note that our assumed surface density profile (\ref{eq:Sigma})-(\ref{eq:Sigma0}) may not fully capture the distribution of $\Sigma$ in real disks. First, pressure forces drive differential precession in a hydrodynamical disk, which can be avoided only under rather special circumstances (Statler 2001).  Second, these equations in their current form do not capture the possible presence of the density waves in the disk driven by the companion perturbation. They can be accounted for by assuming the apsidal angle $\varpi_d$ of the fluid trajectories to vary with the distance in a particular fashion. For simplicity we did not consider such possibility in this work. 

However, even if the expressions (\ref{eq:Sigma})-(\ref{eq:Sigma0}) are only approximate, this does not change our main conclusions about the key role of the disk gravity.  Indeed, we find the values of $A_d$ and $B_d/e_g$, which determine the disk effect on planetesimal dynamics, to not depend sensitively on the power-law indices $p$ and $q$ over a range of reasonable values, see Figure \ref{fig:psi_pl}. Thus, we do not expect our results to change dramatically if the behavior of $\Sigma(a_p)$ and $e_d(a_p)$ were to deviate from the pure power laws in $a_p$.  

Finally, short-term variability of the disk surface density can induce fast-changing torques on planetesimals. These effects cannot be captured by our secular (time-averaged) approach. However, we do not expect them to act coherently on long timescales and therefore to be subdominant for the same reason that the short-period terms of the planetary perturbations play an insignificant role on long time intervals in classical celestial mechanics (Murray \& Dermott 1999).


\subsection{Implications for planetesimal growth}
\label{sect:planet_growth}

Planetesimal growth requires relative velocities of colliding bodies to be small, otherwise they get eroded or destroyed. We use our results to provide some insights on planetesimal accretion in binaries. 

For bodies held together primarily by gravity the threshold collision velocity at which planetesimals can still survive is about the escape speed. Guided by this logic Moriwaki \& Nakagawa (2004) and R13 use the simple criterion 
\ba
e>e_{\rm{crit}} = \frac{2v_{\rm{esc}}}{v_k} & \approx & 6.1 \times 10^{-4} ~\frac{d}{10\mbox{~km}}
\nonumber\\
& \times & \left(\frac{M_\sun}{M_p}\frac{a_p}{1\mbox{~AU}}\frac{\rho}{3\mbox{~g cm}^{-3}}\right)^{1/2}
\label{ecrit}
\ea
as the condition for planetesimal destruction in collisions. Here $v_{\rm{esc}}$ is the escape speed from a planetesimal of a given radius $d$ and $\rho$ is the bulk density of planetesimal material.  

In Figure \ref{fig:ecc_r} we display $e_{\rm{crit}}(a_p)$ by the dashed line and compare it with the characteristic $e_p$ attained by planetesimals as a result of disk+secondary gravitational perturbation for different disk models. One can see that in all models where disk eccentricity $e_d$ is high, comparable to the free-particle diskless eccentricity $e_p^{\rm BB}$ (Figure \ref{fig:ecc_r}a,b), the disk does not help eliminate the fragmentation barrier. This is because $e_p$ cannot drop below $e_d$ and $e_d$ is high.  The situation is clearly more helpful for planetesimal growth in lower-$e_d$ cases, see Figure \ref{fig:ecc_r}c,d, even though it is still not as easy as in the case of axisymmetric disk studied in R13. On the other hand, it has been noted in Rafikov (2013b) that the catastrophic destruction condition (\ref{ecrit}) is likely too conservative and underestimates the ability of planetesimals to survive in mutual collisions. This issue is addressed in more detail in Rafikov \& Silsbee (2014b).

Another potential problem that may arise in low-mass disks with $M_d\sim 10^{-3}M_\odot$ is the presence of secular resonance in the disk, see \S \ref{sect:ecc} and Figure \ref{fig:ecc_r}. There $e_p$ becomes very large in a narrow range of $a_p$, making planetesimal collisions highly destructive. This phenomenon is non-local since high-$e_p$ objects can penetrate other disk regions and destroy planetesimals there as well. 

However, this problem is unlikely to last for a long time as the small number of planetesimals from the vicinity of the secular resonance will be rapidly destroyed in collisions, leaving no more projectiles to destroy the remaining planetesimals in the rest of the disk. Also, our inference of high-$e_p$ at secular resonance is based on linear secular equations (\ref{eq:h_ev})-(\ref{eq:k_ev}), which were derived under the assumption of $e_p\ll 1$, clearly not fulfilled at the resonance. The actual $e_p$ in this part of the disk will be different from our predictions.


\subsection{Lowering planetesimal excitation}
\label{sect:lowering_exc}

Motivated by our results and their implications for planetesimal accretion we next discuss different scenarios (in order of their likely significance) in which relative velocities of planetesimals affected only by the gravity of gaseous disk and binary companion can be considerably lowered.

\subsubsection{Intrinsically low $e_d$}
\label{sect:low_e_d}

The major obstacle for planetesimal growth in high-$e_d$ disks has to do with our general result (\S \ref{sect:ecc}) that $e_p$ is always above the smaller of $e_p^{\rm BB}(a_p)$ and $e_p^{\rm DD}(a_p)\sim e_d$. Thus, one of the most straightforward ways of lowering collision speeds is for the disk to have low $e_d$ either locally or globally for a long period of time. Our current understanding of eccentricity excitation in gaseous disks is based primarily on the results of numerical simulations, which are reviewed in \S \ref{sect:num_compare}. We describe possible ways of lowering $e_d$ there.

\subsubsection{Rapidly precessing disk}
\label{sect:rapid_precess}

When discussing the possibility of disk precession in \S \ref{sect:precession} we noted that the disk-induced contribution to the forced eccentricity can be effectively suppressed if the disk precesses faster than the planetesimals, i.e. if $|\dot\varpi_d|\gg |A|$. In this case $e_p$ can easily be below $e_p^{\rm DD}\sim e_d$. The remaining excitation due to the binary will keep $e_p$ at the level of $e_p^{\rm DB}$, which is low because of the fast planetesimal precession driven by the massive disk, see \S \ref{sect:DB}. Thus, fast disk precession effectively brings planetesimal dynamics to the situation described in R13 and can serve as a mechanism for lowering planetesimal excitation, as long as gas drag can be neglected (Rafikov \& Silsbee 2014a). Whether the gaseous disk can precess at the rate exceeding $|A|$ at separations of several AU, where the giant planets are detected in close binaries, should thus be explored in more detail.

\begin{figure}
\centering
\includegraphics[scale = .45]{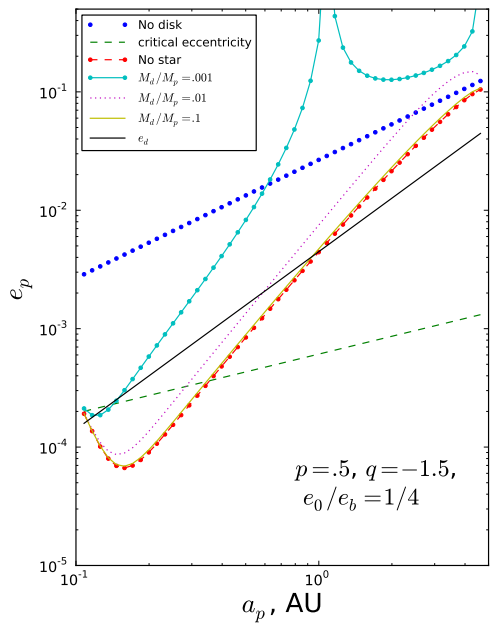}
\caption{Illustration of planetesimal eccentricity behavior for a particular disk model with $p = 0.5$, $q = -1.5$, $e_{0} = 0.05$, extending from 0.1 to 5 AU, in a binary with $a_b=20$ AU, $e_b$ = 0.2, $M_p=M_\odot$, and $\nu=1/2$.  The meaning of the different curves is the same as in Figure \ref{fig:ecc_r}, with the addition of the black line corresponding to $e_d$.  Because this model has $p + q = -1$, the non-axisymmetric part of the disturbing function vanishes (if one neglects edge effects) and $e_p$ is generally quite low, lower than the disk eccentricity $e_d$, and compatible with planetesimal growth (for $\gtrsim 10$ km bodies) for $a_p\lesssim 1$ AU.  We have shown the disk eccentricity $e_d$ (black solid line) to illustrate that the planetesimal eccentricity $e_p$ is much lower than $e_d$ in the inner disk (this is unlike the case of a disk with $p + q \neq -1$).}
\label{fig:low_excit}
\vspace{.1cm}
\end{figure}

\subsubsection{Globally suppressed eccentricity excitation}
\label{sect:global_supp}

Another way of making $e_p^{\rm DD}$ low is hinted to us by equation (\ref{eq:DD_ecc}), which shows that $e_p^{\rm DD}$ can be low even for high $e_d$ if $\psi_2$ is very small. The same is true for $e_p^{\rm BD}$, see equation (\ref{eq:BD_ecc}). This is because the disk then produces zero contribution to the non-axisymmetric component of the disturbing function. Figure \ref{fig:psi_pl} shows that, ignoring the possible edge effects, this is possible e.g. if $p+q=-1$. Our fiducial disk model based on general ideas about the accretion disk physics and their eccentricity excitation has $p=1$ and $q=-1$, which is not compatible with this condition. However, our present understanding of protoplanetary disks in binaries does not allow us to exclude disk models with $p+q=-1$. Interestingly, when $p = 0$ then the disk also has zero contribution to the axisymmetric component; disk with $p=0$ (i.e. uniform disk), $q=-1$ affects neither planetesimal precession nor eccentricity excitation by its gravity in the absence of edge effects.

In Figure \ref{fig:low_excit} we show an example of one such model having $p=0.5$ and $q=-1.5$, with rather high disk eccentricity at the outer disk edge, $e_0=0.25 e_b=0.05$ (for $e_b=0.2$). One can clearly see that in this case $e_p^{\rm DD}$ is low and comparable to $e_{\rm crit}$ at $\sim$AU separations. This should facilitate planetesimal growth on these scales. The disk-induced excitation for this model is not exactly zero due to edge effects. This is more of an issue near the outer edge of the disk because as shown in Appendix \ref{sect:dist_fun} (and illustrated in Figure \ref{fig:pl_phase}) $p + q = -1$ is closer to the line of convergence at the outer edge than at the inner edge\footnote{Asymptotic behavior of equation (\ref{psi2}) shows that for $p+q = -1$, $\psi_2\propto \alpha_2$ as $\alpha_2\to 0$ and $\psi_2\propto \alpha_1^6$ as $\alpha_1\to 0$}.  This means that edge effects are more important in the outer disk for this set of power law indices.  

It is also worth noting that some (though not all) of the lower planetesimal eccentricity in this Figure as compared to Figure \ref{fig:ecc_r} is due to lower assumed disk eccentricity $e_d$ in the inner part of the disk.  However, the drop in $e_p$ as one moves away from the inner edge of the disk reflects the drop in the non-axisymmetric part of the disk disturbing function as the inner edge effect becomes less important and we see that $e_p$ drops well below the local value of $e_d$.

\subsubsection{Locally suppressed eccentricity excitation}
\label{sect:local_supp}

Additionally, there are at least two ways in which $e_p$ can be reduced {\it locally}, within a narrow range of semi-major axes. First, even if the disk does not have $p+q=-1$ {\it globally}, as we assumed in making Figure \ref{fig:low_excit}, there could be parts of the disk in which this condition in fulfilled for a range of $a_p$, for example near the disk edges, where $\Sigma_p$ should be petering out to zero, or near dead zones or opacity transitions, where the material pileup is possible and a non-power law scaling of $\Sigma_p$ is likely. Our results do not directly apply to such situations since we assumed a purely power law behavior of $\Sigma_p(a_p)$ but based on them we can expect that it might be possible to have $\psi_2$ close to zero at radii, near which locally computed $p+q=-\partial \ln\left(e_d\Sigma\right)/\partial\ln a_p$ passes through $-1$ (edge effects mentioned in \S \ref{sect:global_supp} may make situation even more complicated). At this location contributions of the inner and outer disks to $\psi_2$ should nearly cancel each other resulting in low $e_p^{\rm DD}$. Of course, $e_p$ is lowered in this way only if the disk dominates eccentricity excitation, i.e. in the Case DD.

\subsubsection{Favorable disk-binary orientation}
\label{sect:orient}

Second, so far we have always assumed planetesimal eccentricity to be given by the characteristic value $e_{\rm char}$ defined by the equation (\ref{echar}). This approach ignores the dependence of the actual maximum planetesimal eccentricity upon the relative disk-binary orientation, obvious from equation (\ref{eq:e_p}). In particular, in \S \ref{sect:ev_eq} we noted that whenever the conditions (\ref{eq:low_e}) are fulfilled, the maximum eccentricity is much lower than $e_{\rm char}$. Because of the different dependence of $B_d$ and $B_b$ on $a_p$ the first condition can be fulfilled only locally, within a narrow range of radii around $a_p=a_B$ given by equation (\ref{eq:a_B}). Since $a_B$ lies within the disk only for relatively small $M_d$ (see equation \ref{eq:a_B}), we conclude that the first condition is fulfilled only for relatively light disks, $M_d\lesssim 10^{-2}M_\odot$.

For most disk models considered in this work one finds $\psi_2>0$ (see Figure \ref{fig:psi_pl}) and $B_d>0$ (equation \ref{eq:B_d}), while $B_b<0$ (equation \ref{eq:B_b}). Then the second condition in (\ref{eq:low_e}) implies $\varpi_d\approx 0$, i.e. that the binary and the disk apsidal lines need to be {\it aligned} for $e_p$ to be suppressed at $a_B$. For the more atypical cases with $\psi_2<0$ one finds that the disk-secondary {\it anti-alignment} ($\varpi_d\approx \pi$) is necessary to suppress $e_p$ at $a_B$. 

The actual value of $\varpi_d$ for disks inside binaries is not well understood and Okazaki et al. (2002) find numerically that both alignment and anti-alignment are possible for the disks stationary in the binary frame. Needless to say, if the disk is precessing, it is no longer possible for it to be aligned or anti-aligned with the binary companion for a long time and the conditions (\ref{eq:low_e}) are no longer relevant.


\subsection{Comparison with numerical studies}
\label{sect:num_compare}

There exist a number of numerical studies of planetesimal dynamics in binaries which treat structure of the gaseous disk by solving equations of hydrodynamics. However, with the exception of Kley \& Nelson (2007) and Fragner \etal (2011), most of them account only for the effects of gas drag on planetesimal motion and neglect disk gravity. 

The issue of the eccentricity that a gaseous disk develops under the perturbations by the companion has not been settled. Different numerical studies arrive at different conclusions, depending on the physics included in simulations and the numerical methods used. Some simulations find very high values of $e_d$, of order 0.5 at the outer disk edge, that develop if the disk is very extended allowing the operation of an instability related to the $3:1$ resonance studied by Lubow (1991). This mechanism of eccentricity excitation operates even if the companion is on circular orbit. Such a situation is unlikely to apply to the known binary systems, which have relatively massive ($\nu\sim 0.4$) eccentric companions. Circumstellar disks in such systems should be truncated at rather small sizes, excluding the possibility of this instability.

In their SPH study of decretion disks in eccentric Be/X-ray binaries Okazaki et al. (2002) find $e_d\lesssim 0.1$ but the exact value and overall disk behavior (e.g. whether the disk is precessing) strongly depend on the resolution used. Paardekooper et al. (2008) employed a grid-based numerical scheme to simulate a circumstellar disk extending to $0.4a_b$ in a binary with the parameters of the $\gamma$ Cephei system. They find that the value of disk eccentricity very strongly depends on the details of the numerical scheme used, with $e_d(2~\mbox{AU})$ ranging from 0.2 to less than $10^{-2}$. Needless to say this difference should result in very different conclusions regarding the behavior of planetesimals. 

Note that we use $a_{out}=0.25a_b$ in this work, which is smaller that $a_{out}$ used by Paardekooper et al. (2008). A more compact disk is less affected by the binary and might develop smaller $e_d$. At the moment this is just a speculation since the exact value of $a_{out}$ should depend on a number of details such as disk viscosity, binary eccentricity, and so on, see Reg\'aly \etal (2011).

Marzari et al. (2009) find that disk eccentricity is lower when the self-gravity of the disk is properly incorporated in simulations. The same result --- reduction of $e_d$ due to disk self-gravity --- can be seen in circumbinary disks by comparing the study of Pelupessy \& Portegies Zwart (2013), which includes disk self-gravity and finds a regular pattern of low $e_d$, and Marzari et al. (2013), which neglects disk self-gravity and finds very high disk eccentricity. 

This observation is very relevant for our study since we find that massive disks give rise to lower planetesimal eccentricities if disk eccentricity $e_d$ can be reduced below the free-particle eccentricity $e_p^{\rm BB}$, see \S \ref{sect:ecc_behavior}. Lowering $e_d$ by the disk self-gravity would make massive disks even more attractive sites for planetesimal growth. Thus, in line with R13 we suggest that efficiency of planet formation may be a very strong function of the disk mass such that planets form only in binaries with massive disks. Although such systems are rare (Harris et al. 2012) there may be enough of them to explain a handful of known planet-hosting compact binaries.


\section{Summary}
\label{sect:sum}


In this work we explored secular dynamics of planetesimals embedded in an eccentric gaseous disk, with implications for planet formation in binaries. We derived, for the first time, the analytical expression for the disturbing function of a body subject to gravity of a massive, eccentric, confocal and coplanar disk, in the limit when both the disk and planetesimal eccentricities are small (Appendix \ref{sect:dist_fun}). This expression has been used in \S \ref{sect:ev_eq} to understand secular excitation of $e_p$ in presence of both the non-axisymmetric disk and the binary companion. Assuming initially circular orbits and neglecting any dissipation (such as due to gas drag) in this work, we found the general analytical solution for the evolution of planetesimal eccentricity --- equation (\ref{eq:e_p}) --- which shows that $e_p$ oscillates from zero up to some maximum value. 

Both period and amplitude of oscillations depend on properties of the disk and the secondary. Depending on which agent --- disk or secondary --- dominates planetesimal precession and eccentricity excitation, we find four distinct regimes for the $e_p$ behavior. Two of them, in which gravity of eccentric disk dominates planetesimal eccentricity excitation, are novel results of this work. We have shown, in particular, that when the disk dominates both planetesimal precession and eccentricity excitation (so called Case DD, see \S \ref{sect:DD}) characteristic planetesimal eccentricity $e_p$ is of order the local disk eccentricity $e_d$. Thus, the value of $e_d$ sets a lower limit on $e_p$ and essentially determines the characteristic collision speeds of planetesimals. As a result, we generally find that eccentricity of the disk presents a serious obstacle for the growth of planetesimals with sizes of less than several tens of km.  

We then discuss possible ways of lowering $e_p$, which would be favorable for planetesimal growth (\S \ref{sect:lowering_exc}). One of them is for the disk to be massive, typically $\gtrsim 10^{-2}M_\odot$, so that (1) its own self-gravity reduces disk eccentricity $e_d$ as has been suggested by some simulations and (2) disk gravity dominates planetesimal dynamics. Another possibility is for the disk to precess much faster than the precession rate of planetesimal orbits (\S \ref{sect:precession}). Some other ways of lowering $e_p$, both global and local (within a finite range of separations) are also described. These possibilities may represent pathways to planetesimal growth in at least a subset of protoplanetary disks in binary systems.

Despite the neglect of dissipative effects such as gas drag (accounted for in Rafikov \& Silsbee 2014a,b) the present study demonstrates the variety of planetesimal dynamical behaviors driven by the coupled gravitational perturbations of an eccentric disk and the binary. It thus represents an important step in building a complete picture of planetesimal dynamics in binaries. 

Analytical description of the gravitational effects of the eccentric disk derived in this work (Appendix \ref{sect:dist_fun}) can be applied to a variety of other astrophysical problems: planetesimal dynamics in circumbinary disks (Silsbee \& Rafikov, in preparation), dynamics of self-gravitating gaseous and stellar disks, and so on.



\appendix



\section{Disturbing function due to an eccentric disk}
\label{sect:dist_fun}


Here we present a calculation of the disturbing function due to an eccentric disk.  We assume that the disk eccentricity and surface density are given by the power law ansatz (\ref{eq:Sigma0}) and apsidal angle is constant with radius. The latter assumption can be easily relaxed and analytical results obtained for $\varpi_d$ varying as a power law of the semi-major axis of a fluid element. 

There are different ways in which such calculation can be approached. In particular, one can use the analogy with the Gauss averaging method (Murray \& Dermott 1999), which treats the time-averaged potential of a point mass on an eccentric orbit as that produced by an elliptical wire along the orbit with the line density proportional to the time the planet spends at each point of its orbit. In the case of a gaseous disk, we can consider fluid in a narrow elliptical annulus between the two adjacent fluid trajectories. Because of the continuity equation the line density of this fluid along the annulus is also proportional to the time fluid spends at a given location. Given that the density distributions are the same in two cases one can simply employ the expression for the disturbing function given by the Gauss method. For example, secular contribution due to the outer disk becomes
\ba
R^{\rm Gauss}_{out} = n_p^2 a_p^2  \int_{a_p}^{a_{\rm out}} \frac{2 \pi a \Sigma_p(a)}{M_p} \left[\frac{\alpha^2}{8} b_{3/2}^{(1)}\left(\alpha \right) e_p^2 - \frac{\alpha^2}{4} b_{3/2}^{(2)}\left(\alpha \right) e_p e_d(a) \cos{(\varpi_p - \varpi_d)}\right]da,
\label{eq:Gauss}
\ea
where $\alpha=a_p/a$. Similar expression can be written for the inner part of the disk as well. However, both $\int_1 b_{3/2}^{(1)}(\alpha) d \alpha$ and $\int_1 b_{3/2}^{(2)}(\alpha) d\alpha$ are non-convergent, as well as the sum of the inner and outer disk contributions in the vicinity of planetesimal orbit. This is a well-known problem with the Gauss expression for the secular disturbing function (Murray \& Dermott 1999). For this reason we are unable to use Gauss' method to calculate the disturbing function due to an eccentric disk for a planetesimal which is embedded in a disk with no gap.

Instead, we have resorted to a different approach previously used by Heppenheimer (1980) and Ward (1981) to compute the gravitational field of an axisymmetric disk with power law surface density profile. To use this approach for an elliptical disk we had to come up with a number of important modifications. The idea behind this method is to compute the disturbing function directly as
\ba
R({\bf S})=G \left\langle\int\limits_{\bf S} \frac{\Sigma(r_d, \phi_d)r_d dr_d d\phi_d}{\left(r_p^2+r_d^2 - 2r_pr_d\cos\theta\right)^{1/2}}\right\rangle,
\label{eq:R_gen}
\ea
where the integral is taken over the area of the disk ${\bf S}$, angle brackets $\langle ...\rangle$ represent time averaging over planetesimal orbital motion, $r_p$ is the (time-dependent) instantaneous radius of a planetesimal,  $\theta$ is the angle between vectors ${\bf r}_d$ and ${\bf r}_p$, see Figure \ref{fig:geometry}. According to this Figure $\phi_d$ is the polar angle counted from the disk periastron, $\phi_p$ is the angle of the planetesimal with respect to the planetesimal periastron, so that $\theta=\phi_d+\varpi_d-\phi_p-\varpi_p$.

\begin{figure}[ht]
\begin{minipage}[b]{0.45\linewidth}
\centering
\includegraphics[width=\textwidth]{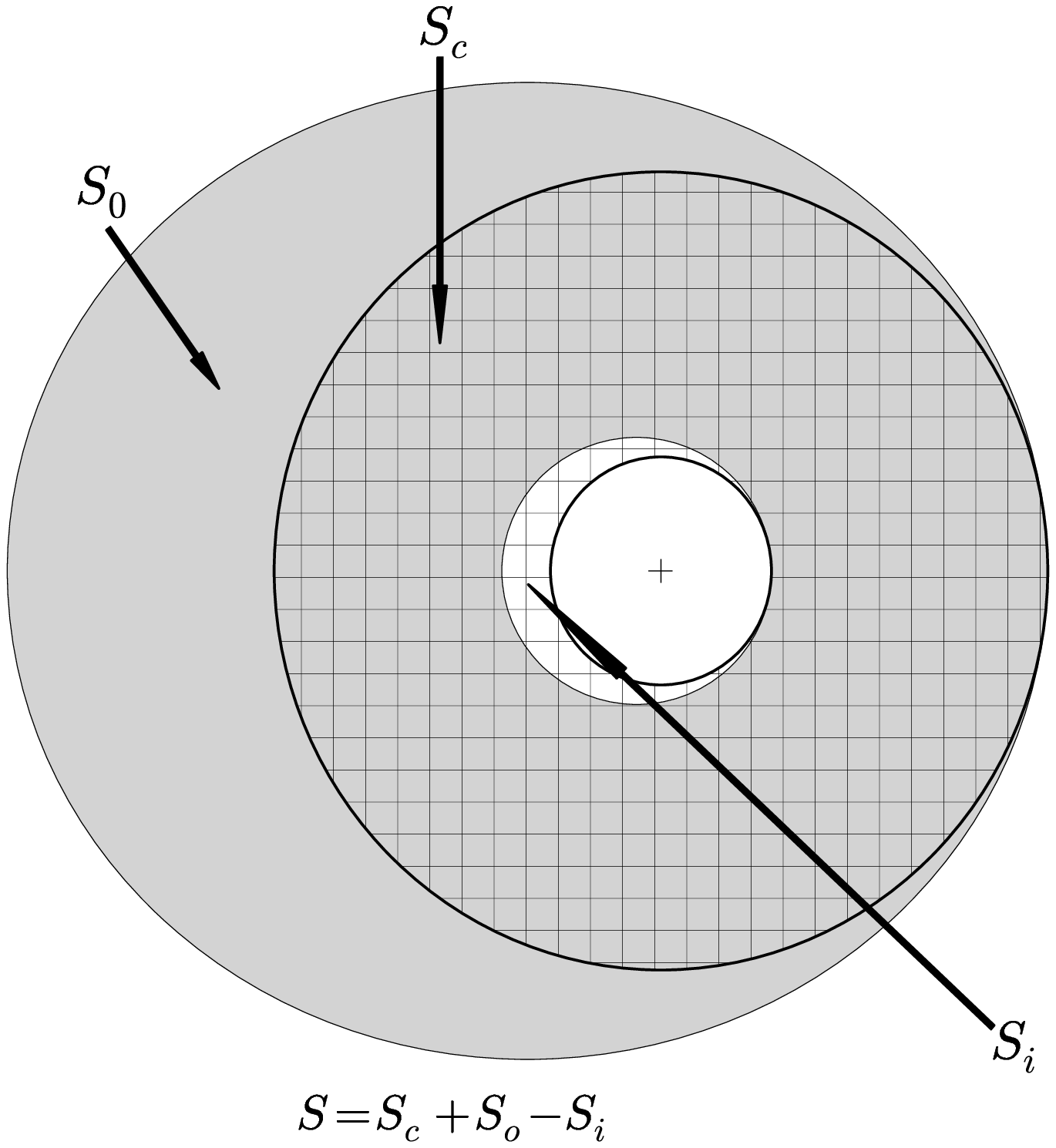}
\caption{Illustration of different integration regions used in calculation of the disk-induced planetesimal disturbing function. The checkered region is the circular annulus ${\bf S}_c$, gray uncheckered crescent is ${\bf S}_o$, white checkered crescent is ${\bf S}_i$. Gray area is the full eccentric disk, ${\bf S}={\bf S}_c+{\bf S}_0-{\bf S}_i$.}
\label{fig:int_regions}
\end{minipage}
\begin{minipage}[b]{0.45\linewidth}
\centering
\includegraphics[width=\textwidth]{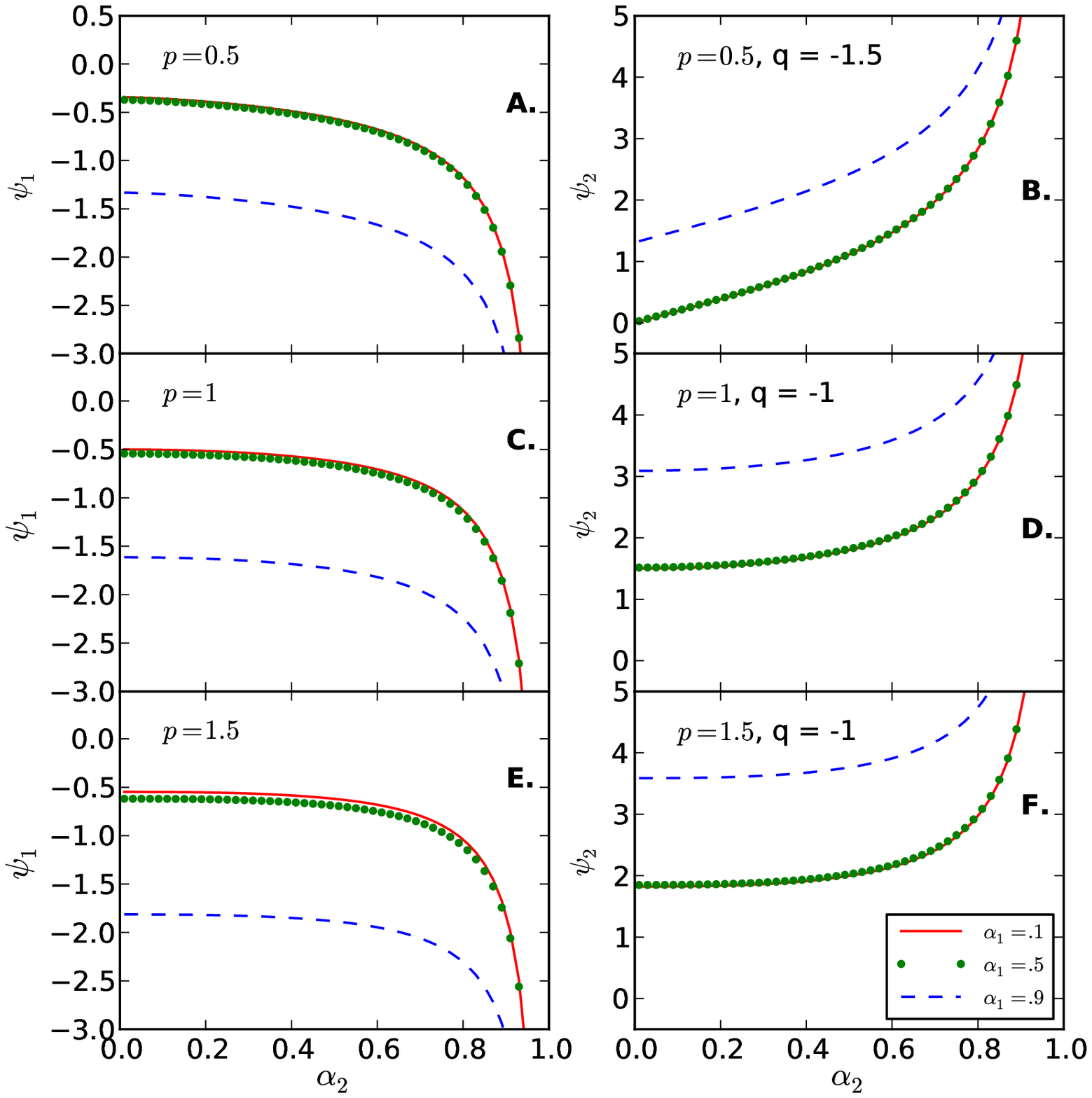}
\caption{Behavior of the pre-factors for the axisymmetric ($\psi_1$) and non-axisymmetric ($\psi_2$) components of the disturbing function near disk edges. Different panels show for different disk models the dependence of $\psi_1$ (left) and $\psi_2$ (right) on $\alpha_2=a_p/a_2$, for different values of $\alpha_1=a_1/a_p$ (shown on panel), with $a_1$ and $a_2$ being the inner and outer semi-major axes of the disk. For the chosen values of $p$ and $q$, $\psi_1$ and $\psi_2$ are essentially constant except as $\alpha_1$ or $\alpha_2$ get close to unity. As a result, $\psi_1$ and $\psi_2$ are essentially constant far from disk edges in these models. This is not the case for the model with $p+q=-1$ in panel b, which is featured in \S \ref{sect:global_supp} and Figure \ref{fig:low_excit}.}
\label{fig:psi_alpha}
\end{minipage}
\end{figure}

We divide the disk up into three regions as shown in Figure \ref{fig:int_regions}, so that ${\bf S}={\bf S}_c+{\bf S}_0-{\bf S}_i$. Here  ${\bf S}_c$ is the annulus bounded by circles with radii equal to the periastron of the outer disk edge $a_{\rm out}[1-e_d(a_{\rm out})]$ and the periastron of the inner disk edge $a_{\rm in}[1-e_d(a_{\rm in})]$; ${\bf S}_o$ is the outer crescent region bounded the outer circle of ${\bf S}_c$ on the inside and the outermost elliptical trajectory on the outside; ${\bf S}_i$ is the inner crescent region bounded the inner circle of ${\bf S}_c$ on the outside and the innermost elliptical trajectory on the inside. The full disturbing function of an eccentric disk is given by 
\begin{equation}
R({\bf S}) = R({\bf S}_c) + R({\bf S}_o) - R({\bf S}_i)
\label{eq:disturb_split}
\end{equation}
We now separately calculate the contributions due to different regions using an extension of the method employed by Heppenheimer (1980).

\subsection{Contribution from the annular region ${\bf S}_c$}
\label{eq:S_c}

We start by calculating the contribution from the annular region ${\bf S}_c$. In the following we define for brevity $a_{\rm in}=a_1$, $a_{\rm out}=a_2$, $e_d(a_{\rm in})=e_1$, $e_d(a_{\rm out})=e_2$, with $r_{d,{\rm in}}=a_1(1-e_1)$ and $r_{d,{\rm out}}=a_2(1-e_2)$ being the inner and outer radii of ${\bf S}_c$. We can write 
\ba
R({\bf S}_c)=\Bigg \langle G \int\limits^{r_{d,{\rm out}}}_{r_{d,{\rm in}}} r_d dr_d \int\limits_{0}^{2\pi} \frac{\Sigma(r_d, \phi_d)}{\sqrt{r_p^2+r_d^2 - 2r_pr_d\cos\theta}}d\theta \Bigg \rangle
\label{eq:S_c_gen}
\ea 
where we have used the fact that $d\phi_d=d\theta$. 

As given in Statler (2001) and using equation (\ref{eq:Sigma}), to first order in $e_d$,
\begin{equation}
\Sigma(r_d, \phi_d) = \Sigma_p(r_d)+ e_d\left[\zeta(r_d)\Sigma_p(r_d)\left(\cos\phi_d - 1\right) + r_d \cos\phi_d \frac{d\Sigma_p(r_d)}{dr_d}\right]
\label{eq:lin_ex}
\end{equation}
where $\zeta(r_d)$ was defined after equation (\ref{eq:Sigma}). Note that $\Sigma_p$ is considered to be a function of the {\it semi-major axis} of a fluid element passing through a given point in the disk, as stated after equation (\ref{eq:Sigma}). For that reason $\Sigma(r_d,0)\neq \Sigma_p(r_d)$ but $\Sigma(r_d,0)= \Sigma_p(r_d/(1-e_d))$ (to second order in $e_d$), i.e. at the semi-major axis $r_d/(1-e_d)$ for which $r_d$ is the periastron distance. 

Classical secular theory neglects terms in the disturbing function which are higher order than $e_p^2$ in planetesimal eccentricity, see \S \ref{sect:dist_disk}. Thus, in our subsequent calculations we will retain only terms proportional to $e_p^2$ and $e_pe_d$; terms of higher order in $e_d$ are neglected because, by assumption, $e_d\ll 1$.  As we will see below, terms with no $\phi_d$ dependence lead to corrections of order $e_p^2$.  Therefore we can drop the terms with no $\phi_d$ dependence which are also proportional to $e_d$.

With this in mind, we write the contribution to the disturbing function from the annular component ${\bf S}_c$ as
\ba
R({\bf S}_c) & = & I_1 + I_2,
\label{S_c_separ}\\
I_1 & = &  \left \langle G \int\limits_{r_{d,{\rm in}}}^{r_{d,{\rm out}}} dr_d~ \Sigma_p(r_d) r_d \int\limits_{0}^{2\pi}  \frac{d\theta}{\sqrt{r_p^2+r_d^2-2r_pr_d\cos{\theta}}} \right \rangle
\label{I1}\\
I_2 & = &  \left \langle G \int\limits_{r_{d,{\rm in}}}^{r_{d,{\rm out}}} dr_d \left[\zeta(r_d)\Sigma_p(r_d) + r_d \frac{d\Sigma_p(r_d)}{dr_d}\right]e_d(r_d) r_d   \int\limits_{0}^{2\pi} \frac{\cos{(\theta + v)} d\theta}{\sqrt{r_p^2+r_d^2-2r_pr_d\cos{\theta}}} \right \rangle.
\label{I2}
\ea
Here we expressed $\phi_d=\theta + v$, where $v = \varpi_p -\varpi_d + \phi_p$, see Figure \ref{fig:geometry}. We now evaluate these two contributions.


\smallskip
\smallskip
{\bf Evaluation of $I_1$}
\smallskip
\smallskip

From the definition (\ref{eq:Laplace}) of the Laplace coefficients we can write the inner integral over $\theta$ in equation (\ref{I1}) as $(\pi/r_d)b_{1/2}^{(0)}(r_p/r_d)$ for $r_p<r_d$ (outer disk) and $(\pi/r_p)b_{1/2}^{(0)}(r_d/r_p)$ for $r_p>r_d$ (inner disk). Assuming surface density prescription (\ref{eq:Sigma0}) we can write 
\ba
I_1 & = & \pi G \Sigma_0\left \langle \int\limits_{r_{d,{\rm in}}}^{r_p} \left(\frac{a_{\rm out}}{r_d}\right)^p \frac{r_d}{r_p} b_{1/2}^{(0)}\left(\frac{r_d}{r_p}\right)dr_d+ 
\int\limits_{r_p}^{r_{d,{\rm out}}} \left(\frac{a_{\rm out}}{r_d}\right)^p b_{1/2}^{(0)}\left(\frac{r_p}{r_d}\right)dr_d \right\rangle.
\label{I1_1}
\ea
We now define auxiliary function 
\begin{equation}
\label{I}
I(x, y, z) \equiv \int_x^1 \alpha^y b_{1/2}^{(z)}(\alpha) d\alpha,
\end{equation}
and a new constant factor
\begin{equation}
\label{c1}
K = \pi G \Sigma_0 a_{\rm out}^p a_p^{1-p}.
\end{equation}
With these definitions we re-write expression (\ref{I1_1}) as
\begin{equation}
I_1 = K\left\langle \left(\frac{r_p}{a_p}\right)^{1-p}\left[I(a_1/r_p, 1-p, 0) + I(r_p/a_2, p-2, 0) \right] \right\rangle.
\label{eqpotalphaprime}
\end{equation}
We note that $a_1$ and $a_2$ in these expressions approximate $r_{d,in}=a_1(1-e_1)$ and $r_{d,out}=a_2(1-e_2)$, correspondingly. However, the difference is a correction linear in disk eccentricity $e_d$ and should be ignored for $I_1$. 

We now proceed to the last, time averaging, step. For illustration we perform it first on the second integral in this expression, by expanding it in Taylor series in small quantity $r_2 - \alpha_2$, where $r_2 = r_p/a_2$, and $\alpha_2 = a_p/a_2$. We have
\begin{equation}
\label{eqpotalpha}
I(r_p/a_2, p-2, 0) = I(\alpha_2, p-2, 0) - (r_2 - \alpha_2) \alpha_2^{p-2} b_{1/2}^{(0)}(\alpha_2) - \frac{(r_2 - \alpha_2)^2}{2} \frac{d}{d\alpha_2}\left[\alpha_2^{p-2} b_{1/2}^{(0)}(\alpha_2)\right].
\end{equation}
We may relate $r_p$ and $a_p$ using the eccentric anomaly $E$ as $r_p = a_p(1-e_p\cos E)$. Then $r_2 - \alpha_2 = -\alpha_2 e_p \cos E $ and 
\begin{equation}
\label{helperOne}
\left(\frac{r_p}{a_p}\right)^{1-p} = 1 - (1-p)e_p \cos E - \frac{p(1-p)}{2} e_p^2\cos^2 E.
\end{equation}
Using these relations, the second integrand in (\ref{eqpotalphaprime}) becomes (retaining only terms up to $e_p^2$)
\ba
&& K \left\langle 1 - (1-p)e_p \cos E - \frac{p(1-p)}{2} e_p^2\cos^2 E \right\rangle I(\alpha_2, p-2, 0)
\nonumber  \\
+ && K \left\langle \left[1 - (1-p)e_p \cos E \right]  e_p \cos E ~\alpha_2^{p - 1} b_{1/2}^{(0)}(\alpha_2)\right\rangle 
-\frac{K}{2} \left\langle \alpha_2^2 e_p^2 \cos^2 E \frac{\partial}{\partial \alpha_2} \left[\alpha_2^{p - 2} b_{1/2}^{(0)}(\alpha_2)\right] \right\rangle.
\label{eq:interm}
\ea
Using $\langle \cos E \rangle = -e_p/2$ and $\langle \cos^2 E  \rangle = 1/2$, equation (\ref{eq:interm}) reduces to
\ba
\label{eqpotans1}
K\left[(1+\frac{e_p^2}{4}(1-p)(2-p) \right] I(\alpha_2, p-2, 0)+
K\frac{e_p^2}{4}\left[2(p-1)\alpha_2^{p-1}b_{1/2}^{(0)}(\alpha_2)  
- \frac{d}{d\alpha_2} \left[\alpha_2^p b_{1/2}^{(0)}(\alpha_2)\right] \right].
\ea
We can apply the identical procedure to the first integral of equation (\ref{eqpotalphaprime}), resulting in  
\ba
\label{eqpotans2}
  K\left[1+\frac{e_p^2}{4}(1-p)(2-p) \right] I(\alpha_1, 1-p, 0)
+K\frac{e_p^2}{4}\left[2(2-p) \alpha_1^{2-p} b_{1/2}^{(0)}(\alpha_1) - \frac{\partial}{\partial \alpha_1}\left[\alpha_1^{3-p}b_{1/2}^{(0)}(\alpha_1)\right]\right],
\ea
where $\alpha_1 = a_1/a_p$.  Note that the term second order in $e_p$ must be included in $r_1 - \alpha_1 = \alpha_1 e_p \cos{E} + \alpha_1 e_p^2 \cos^2{E}$, where $r_1=a_1/r_p$.  The sum of (\ref{eqpotans1}) and (\ref{eqpotans2}) is equal to $I_1$ and represents the axisymmetric part of the disk disturbing function from the region ${\bf S}_c$.


\smallskip
\smallskip
{\bf Calculation of $I_2$}
\smallskip
\smallskip

In order to calculate $I_2$ --- the non-axisymmetric component of $R({\bf S}_c)$ we use the prescription (\ref{eq:Sigma0}) for $\Sigma_p$ and $e_d$, and expand $\cos(\theta+v)$:
\ba
I_2 & = &  -G(p+q)\left \langle \int\limits_{r_{d,{\rm in}}}^{r_{d,{\rm out}}} dr_d ~ r_d \Sigma_0 e_0\left(\frac{a_{\rm out}}{r_d}\right)^{p+q}  \int\limits_{0}^{2\pi} \frac{\cos{\theta}\cos{v} - \sin{\theta} \sin{v}}{\sqrt{r_p^2+r_d^2-2r_pr_d\cos{\theta}}} d\theta \right \rangle.
\ea
In the inner integral over $\theta$ terms with $\sin\theta$ in the numerator vanish upon integration, while the terms with $\cos\theta$ result in Laplace coefficients $b_{1/2}^{(1)}$, see definition (\ref{eq:Laplace}). Separately accounting for the contributions from the inner and outer disks when integrating over $r_d$ we obtain, analogous to equation (\ref{eqpotalphaprime})
\ba
\label{nonaxisymm}
I_2 = -Ke_d(a_p)(p + q)\left \langle  \cos{v}\left(\frac{r_p}{a_p}\right)^{1-p-q}\left[ I(r_p/a_2, p + q - 2, 1)+I(a_1/r_p, 1-p-q, 1) \right] \right \rangle.
\ea

The final step of time-averaging is somewhat more challenging here because the $\cos v$ term introduces additional time-dependence through $\phi_p$. It can be taken care of using the definition $v = (\varpi_p - \varpi_d) + \phi_p$ and the relation $\phi_p = E + e_p\sin E$ accurate to linear order in $e_p$. As before, we also expand integrals in (\ref{nonaxisymm}) in a series in the small quantities $r_{1,2} - \alpha_{1,2}$. Since we are not interested in the terms $O\left(e_d e_p^2\right)$ and higher order (small factor $e_d$ is already present in equation (\ref{nonaxisymm})), we only expand to first order in $e_p$. As a result of tedious but straightforward calculation we find
\ba
I_2= -Ke_d(a_p)e_p\cos{(\varpi_p - \varpi_d)}(p + q) && \bigg[\frac{(p+q-3)}{2}I(\alpha_1, 1-p - q, 1) - \frac{1}{2} \alpha_1^{2-p-q}b_{1/2}^{(1)}(\alpha_1)
\nonumber \\
&&
+ \frac{(p+q-3)}{2}I(\alpha_2, p + q - 2, 1) + \frac{1}{2} \alpha_2^{p+q-1}b_{1/2}^{(1)}(\alpha_2)\bigg].
\label{eqpotans3}
\ea 
This completes our calculation of $R({\bf S}_c)$.


\subsection{Contribution from the inner crescent ${\bf S}_i$}

We now calculate the disturbing function $R({\bf S}_i)$, given by equation (\ref{eq:R_gen}) with integration carried out over the inner crescent ${\bf S}_i$. The width of the crescent is $O(e_d)$ meaning that we need to keep all variables only up to first order in $e_p$. The integrand, which led to an axisymmetric contribution in the case of $R({\bf S}_c)$ now leads to a non-axisymmetric contribution when integrated over this non-axisymmetric region of the disk.

Consider an ellipse with periastron distance $a_{p,1}=a_1(1-e_1)$ and apoastron distance $a_{a,1}=a_1(1-e_1)$ bounding ${\bf S}_i$ on the outside. Define the angle $\xi(r_d)$ as the angle between the periastron of this ellipse and the point of intersection of the ellipse and a circle of radius $r_d$, $a_{p,1}<r_d<a_{a,1}$. Linearizing the equation of an ellipse $r_d = a_1(1-e_1^2)/\left[1+e_1\cos\xi(r_d)\right]$ in $e_1$ we get $r_d = a_1(1-e_1\cos\xi(r_d))$. This yields 
\begin{equation}
\xi(r_d)  = \arccos\frac{a_1-r_d}{e_1a_1},
\label{eq:xi}
\end{equation}
where the arccos function is the inverse cosine function. We write explicitly
\begin{equation}
R({\bf S}_i) = \left \langle G \Sigma_0 a_{\rm out}^{p} \int\limits_{a_{p,1}}^{a_{a,1}} \frac{r_d^{1-p}}{r_p} dr_d \int\limits_{\xi-\Delta\varpi - \phi_p}^{2\pi-\xi-\Delta\varpi - \phi_p} \frac{d\theta}{\sqrt{1+ \alpha^{\prime 2} - 2\alpha^\prime \cos\theta}} \right \rangle
\end{equation}
where $\alpha^\prime = r_d/r_p\approx a_1/r_p$, and $\Delta\varpi = \varpi_p - \varpi_d$. Then using the relation
\begin{equation}
\left(1 + \alpha^{\prime 2} - 2\alpha^\prime \cos{\theta}\right)^{-1/2} = \frac{1}{2} b_{1/2}^{(0)}(\alpha^\prime) + \sum_{j = 1}^\infty b_{1/2}^{(j)}(\alpha^\prime) \cos{(j \theta)}
\end{equation}
the inner integral over $\theta$ becomes
\begin{equation}
\left[\pi - \xi(r_d)\right] b_{1/2}^{(0)}(\alpha^\prime) - \sum_{j = 1}^\infty \frac{2}{j} b_{1/2}^{(j)}(\alpha^\prime) \sin\left[j\xi(r_d)\right]\cos\left[j(\Delta\varpi + \phi_p)\right].
\end{equation}
Then we may write 
\ba
R({\bf S}_i) & = & G \Sigma_0 a_{\rm out}^p \Bigg \langle r_p^{-1}\int\limits_{a_{p,1}}^{a_{a,1}}  r_d^{1-p} dr_d \bigg[b_{1/2}^{(0)}\left(\frac{a_1}{r_p}\right)\left[\pi - \xi(r_d)\right] 
\nonumber\\
& - & \sum_{j = 1}^\infty \frac{2}{j} b_{1/2}^{(j)}\left(\frac{a_1}{r_p}\right) \sin\left[j\xi(r_d)\right] \Big[\cos{(j\Delta\varpi)}\cos{(j\phi_p)} - \sin{(j\Delta\varpi)} \sin{(j\phi_p)}\Big] \bigg] \Bigg \rangle
\label{eq:interm2}
\ea
We will use the following definite integrals
\ba
\int\limits_{a_{p,1}}^{a_{a,1}} \Big[\pi - \xi(r_d)\Big]dr_d = \pi a_1 e_1, 
\quad\quad\quad\quad 
\int\limits_{a_{p,1}}^{a_{a,1}} \sin\left[\xi(r_d)\right] dr_d = \frac{\pi}{2} e_1 a_1,
\quad\quad\quad\quad 
\int\limits_{a_{p,1}}^{a_{a,1}} \sin\left[j\xi(r_d)\right] dr_d = 0
\ea
for integer $j>1$. Then in (\ref{eq:interm2}) we may ignore the terms in the sum with $j>1$:
\begin{equation}
R({\bf S}_i) =\pi G \Sigma_0 \left(\frac{a_{\rm out}}{a_1}\right)^p e_1 a_1^2 \left \langle r_p^{-1} \left[b_{1/2}^{(0)}\left(\frac{a_1}{r_p}\right) - b_{1/2}^{(1)}\left(\frac{a_1}{r_p}\right)\left[\cos{\Delta\varpi} \cos{\phi_p} - \sin{\Delta\varpi} \sin{\phi_p}\right] \right] \right \rangle
\end{equation}
where $e_1$ and $a_1$ are the disk eccentricity and semi-major axis respectively, evaluated at the inner edge. Using $a_1/r_p = \alpha_1 + e_p \alpha_1 \cos E$ and the relation between $\phi_p$ and $E$ this becomes
\ba
 R({\bf S}_i) & = & \pi G \Sigma_0 \left(\frac{a_{\rm out}}{a_1}\right)^p a_1 e_1 \Bigg \langle\left(\alpha_1 + e_p \alpha_1 \cos{E}\right) \Bigg\{ b_{1/2}^{(0)}(\alpha_1) + e_p \alpha_1 \cos{E} \frac{ \partial b_{1/2}^{(0)}}{\partial \alpha_1} -  \left[b_{1/2}^{(1)}(\alpha_1) + e_p \alpha_1 \cos{E} \frac{ \partial b_{1/2}^{(1)}}{\partial \alpha_1}\right]
\nonumber \\
& \times & \left(\cos\Delta\varpi \cos E - \sin \Delta \varpi \sin E- e_p \cos{\Delta\varpi}\sin^2{E} - e_p \sin{\Delta\varpi} \cos{E} \sin{E} \right)\Bigg\}  \Bigg \rangle.
\label{SiGiantEquation}
\ea
Expanding all products in this expression one gets a total of 20 terms. It is straightforward to angle-average them as before. Keeping only terms of order $O(e_1e_p)$ and substituting $e_1 = e_0\left(a_{\rm out}/a_1\right)^q$ we find that the disturbing function from the inner crescent is given by
\ba
R({\bf S}_i) =\frac{1}{2} K e_d(a_p)e_p \cos\left(\varpi_p - \varpi_d\right)~  \alpha_1^{2-p - q}\left[b_{1/2}^{(1)}(\alpha_1) - \alpha_1\frac{\partial b_{1/2}^{(1)}}{\partial \alpha_1}\right].
\label{eq:R_S_i}
\ea


\subsection{Contribution from the outer crescent ${\bf S}_o$}

The derivation of $R({\bf S}_o)$ follows the same basic concept as that of $R({\bf S}_i)$ except that now $\alpha^\prime = r_p/r_d$. As a result one finds the contribution of the outer crescent ${\bf S}_o$ to be given by 
\begin{equation}
R({\bf S}_o)=\frac{1}{2} K e_d(a_p)e_p\cos\left(\varpi_p - \varpi_d\right)~
\alpha_2^{p + q - 1}\left[2 b_{1/2}^{(1)}(\alpha_2) + \alpha_2 \frac{\partial b_{1/2}^{(1)}}{\partial \alpha_2}\right].
\label{eq:R_S_0}
\end{equation}


\subsection{Putting everything together.}

Plugging equations (\ref{eqpotans1}), (\ref{eqpotans2}), (\ref{eqpotans3}), (\ref{eq:R_S_i}), (\ref{eq:R_S_0}) into the expression (\ref{eq:disturb_split}) we find that the total eccentric disk-induced disturbing function, including the eccentricity independent term and terms proportional to $e_p^2$ and $e_de_p$, is given by
\begin{equation}
R = K\left[\psi_0 + \psi_1 e_p^2 + \psi_2 e_d(a_p)e_p \cos(\varpi_p - \varpi_d)\right]
\label{eq:dist_f}
\end{equation}
with 
\ba
\psi_0(\alpha_1,\alpha_2) & = & I(\alpha_1, 1-p, 0) + I(\alpha_2, p - 2, 0)
\label{eqpotc0}\\
\psi_1(\alpha_1,\alpha_2)  & = & \frac{1}{4}\left[(1-p)(2-p) \psi_0(\alpha_1,\alpha_2) + 2(p-1)\alpha_2^{p-1}b_{1/2}^{(0)}(\alpha_2)  - \frac{d}{d\alpha_2} \left(\alpha_2^p b_{1/2}^{(0)}(\alpha_2)\right)\right. 
\nonumber \\
& + & \left. 2(2-p) \alpha_1^{2-p} b_{1/2}^{(0)}(\alpha_1) - \frac{\partial}{\partial \alpha_1}\left(\alpha_1^{3-p}b_{1/2}^{(0)}(\alpha_1)\right)\right]
\label{psi1}\\
\psi_2(\alpha_1,\alpha_2) & = & -\frac{(p + q)(p+q-3)}{2} 
\left[I(\alpha_1, 1-p - q, 1) +  I(\alpha_2, p + q - 2, 1)\right]
\nonumber \\
& + &  
\frac{\alpha_1^{2 - p - q}}{2} \left[\left(p+q-1\right)b_{1/2}^{(1)}(\alpha_1) + \alpha_1\frac{\partial b_{1/2}^{(1)}}{\partial \alpha_1}\right] + 
\frac{\alpha_2^{p + q - 1}}{2} \left[\left(2-p-q\right) b_{1/2}^{(1)}(\alpha_2) + \alpha_2 \frac{\partial b_{1/2}^{(1)}}{\partial \alpha}\right]
\label{psi2}
\ea
This completes our calculation of the disturbing function due to an eccentric disk with properties given by equation (\ref{eq:Sigma0}).


\subsection{Asymptotic behavior}

Astrophysical disks typically span several orders of magnitude in radius. It is then plausible that far from the disk boundaries we can ignore the edge effects, i.e. the expression for the disturbing function does not depend on the $a_{\rm in}$ and $a_{\rm out}$ as $a_{\rm in}\to 0$ and $a_{\rm out}\to \infty$. This corresponds to the limit of $\alpha_{1,2}\to 0$. Using Taylor expansion 
\ba
b_{1/2}^{(0)}(\alpha) = 2 + \frac{\alpha^2}{2},~~~~~~
b_{1/2}^{(1)}(\alpha) = \alpha + \frac{3}{8} \alpha^3,
\ea
for small $\alpha$ in equations (\ref{psi1})-(\ref{psi2}), we determined that $\psi_1$ is convergent and independent of $\alpha_{1,2}$ as $\alpha_{1,2}\to 0$ go to zero for $-1<p < 4$. Similarly, $\psi_2$ is convergent as $\alpha_{1,2}\to 0$ for $-2 < p + q < 5$. Convergence limits are illustrated in Figure \ref{fig:pl_phase}.

Provided that the disturbing function is dominated by the local parts of the disk (i.e. the values of $p$ and $q$ fall within the white region in Figure \ref{fig:pl_phase}) and the values of coefficients $\psi_{1,2}$ are independent of $\alpha_{1,2}$ when the disk edges are well separated from the planetesimal semi-major axis ($\alpha_{1,2}\to 0$), we can obtain simpler analytical expressions for these coefficients. Indeed, using the fact that $b_{1/2}^{(0)}(\alpha)=(4/\pi){\bf K}(\alpha)$, $b_{1/2}^{(1)}(\alpha)=(4/\pi\alpha)\left[{\bf K}(\alpha)-{\bf E}(\alpha)\right]$ (here ${\bf E}$ and ${\bf K}$ are complete elliptic integrals) and series expansions (Gradshteyn \& Ryzhik 1994)
\ba
{\bf K}(\alpha)=\frac{\pi}{2}\left(1+\sum\limits_{n=1}^{\infty}A_n\alpha^{2n}\right),~~~
{\bf E}(\alpha)=\frac{\pi}{2}\left(1-\sum\limits_{n=1}^{\infty}\frac{A_n}{2n-1}\alpha^{2n}\right),~~~A_n=\left[\frac{(2n)!}{2^{2n}(n!)^2}\right]^2,
\label{eq:ellipt}
\ea
we can provide asymptotic expressions for $\psi_{1,2}$ as follows:
\ba
\psi_1(0,0) & = & -\frac{1}{2}+\frac{(1-p)(2-p)}{2}\sum\limits_{n=1}^{\infty}\frac{(4n+1)A_n}{(2n+2-p)(2n+p-1)},
\label{eq:psi1_as}\\
\psi_2(0,0) & \to & \frac{3}{2}-(p+q)(p+q-3)
\sum\limits_{n=2}^{\infty}\frac{2n(4n-1)A_n}{(2n-1)(2n+1-p-q)(2n-2+p+q)}.
\label{eq:psi2_as}
\ea
The behavior of $\psi_1(0,0)$ and $\psi_2(0,0)$ as functions of $p$ and $p+q$ respectively are shown in Figure \ref{fig:psi_pl}.


\section{Details of the numerical verification}
\label{ap:num_ver}

Here we describe the details of the numerical verification of our analytical results, see \S \ref{sect:compare}. We directly integrated orbits of planetesimals affected by the gravity of an eccentric disk using the MERCURY package (Chambers 1999). All our integrations employed Bulirsch-Stoer algorithm (Press \etal 1992). Accelerations due to the gravity of an eccentric disk ${\bf g}_d$ (used as an input for our integrations) were computed at different positions and for different disk parameters via direct numerical integration as
\ba
{\bf g}_d({\bf r})=-G\int\limits_{\bf S}\Sigma({\bf r}_d) 
\frac{{\bf r}-{\bf r}_d}{|{\bf r}-{\bf r}_d|^3}
d{\bf S}({\bf r}_d),
\label{eq:accel}
\ea
where $d{\bf S}({\bf r}_d)$ is a surface element centered on ${\bf r}_d$. This two-dimensional integral was performed using standard integration by quadratures in SciPy.  We used a small softening parameter in the integrand to better handle the singularity, and verified convergence to within a percent as we lowered the value of this parameter.  The surface density of the eccentric disk was assumed to be given directly by equation (\ref{eq:Sigma}). In this calculation we did not make an assumption of $e_d\ll 1$ and thus were not expanding equation (\ref{eq:Sigma}) in powers of $e_p$ (as opposed to equation (\ref{eq:lin_ex})).


\section{Precessing disk}
\label{sect:prec_App}
 

Here we explore secular evolution of planetesimals in the case of a disk precessing according to a simple linear prescription $\varpi_d(t)=\varpi_{d0}+\dot\varpi_d t$. Plugging it into Lagrange equations (\ref{eq:h_ev})-(\ref{eq:k_ev}) one finds the following solution for the components of eccentricity vector $(k_p,h_p)$ with the initial conditions $k_p(0) = 0$, $h_p(0) = 0$:
\ba
k_p(t) & = & \frac{1}{A(A-\dot \varpi_d)} \left\{AB_d\left[\cos(At+\varpi_{d0}) - \cos(\dot \varpi_d t + \varpi_{d0})\right] + B_b(A-\dot \varpi_d)\left[\cos{(At)} - 1\right]\right\}.
\label{eq:k1}\\
h_p(t) & = & \frac{1}{A(A-\dot \varpi_d)} \left\{AB_d\left[\sin{(At+\varpi_{d0})} - \sin{(\dot \varpi_d t + \varpi_{d0})}\right] + B_b(A-\dot \varpi_d)\sin{(At)}\right\},
\label{eq:h1}
\ea
which generalizes solution (\ref{eq:decompose}) to the case of non-zero precession. It can again be written as the sum of three distinct contributions as described in \S \ref{sect:precession}:
\ba
{\bf e}_p(t)=\left\{
\begin{array}{l}
k_p(t)\\
h_p(t)
\end{array}
\right\} & = & 
-\frac{B_b}{A}
\left\{
\begin{array}{l}
1 \\
0
\end{array}
\right\}
-\frac{B_d}{A-\dot \varpi_d}
\left\{
\begin{array}{l}
\cos\varpi_d(t) \\
\sin\varpi_d(t)
\end{array}
\right\}
\nonumber\\
& + & \frac{\left[(AB_d)^2 + \left(B_b(A-\dot \varpi_d)\right)^2 + 2AB_dB_b(A - \dot \varpi_d)\cos\varpi_{d0}\right]^{1/2}}{A(A - \dot \varpi_d)}
\left\{
\begin{array}{l}
\cos (At + \phi) \\
\sin (At + \phi)
\end{array}
\right\}
\label{eq:e_precess}
\ea
where $\phi$ is a phase defined analogous to (\ref{eq:phase}) and is a function of $\varpi_{d0}$, $A$, $B_d$ and $B_b$. Note that in the case of a precessing disk, forced eccentricity due to the disk changes in time.

\end{document}